\documentclass[aps,prl,reprint,groupedaddress]{revtex4-1}

\usepackage{graphicx}
\usepackage{color}
\usepackage{amsmath}
\usepackage{amsfonts}
\usepackage{amssymb}
\usepackage{amsthm}

\DeclareMathOperator{\maximize}{maximize}

\DeclareMathOperator{\subjectto}{subject\;to}
\DeclareMathOperator{\real}{Re}
\DeclareMathOperator{\imag}{Im}
\DeclareMathOperator{\Ei}{Ei}

\begin{document}

\title{On the Fundamental Limitations of Imaging with Evanescent Waves}
\author{Alexander Y. Piggott}
\author{Logan Su}
\author{Jan Petykiewicz}
\author{Jelena Vu\v{c}kovi\'{c}}
\affiliation{Ginzton Laboratory, Stanford University, Stanford, California 94305, USA}
\date{\today}

\begin{abstract}
There has been significant interest in imaging and focusing schemes that use evanescent waves to beat the diffraction limit, such as those employing negative refractive index materials or hyperbolic metamaterials. The fundamental issue with all such schemes is that the evanescent waves quickly decay between the imaging system and sample, leading to extremely weak field strengths. Using an entropic definition of spot size which remains well defined for arbitrary beam profiles, we derive rigorous bounds on this evanescent decay. In particular, we show that the decay length is only $w / \pi e \approx 0.12 w$, where $w$ is the spot width in the focal plane, or $\sqrt{A} / 2 e \sqrt{\pi} \approx 0.10 \sqrt{A}$, where $A$ is the spot area. Practical evanescent imaging schemes will thus most likely be limited to focal distances less than or equal to the spot width.
\end{abstract}

\maketitle
Traditional optical microscopes are limited to a resolution of approximately half a wavelength due to the diffraction of light. In the past few decades, there has been great deal of interest in pushing the resolution of optical microscopes beyond this limit by using evanescent waves. This was first achieved by near-field scanning optical microscopy (NSOM), which uses a sharp metal tip or aperture to tightly focus light. The NSOM approach has significant drawbacks since the tip or aperture must first be brought into extremely close proximity with the sample, and then scanned across the specimen line-by-line to produce an image \cite{rcdunn_cr1999}. To circumvent the limitations of NSOM, a variety of metamaterial-based approaches have been proposed, which use carefully-engineered metamaterials to amplify or otherwise generate evanescent fields over a broad area. This approach was pioneered by Pendry in 2000, who proposed that a slab of negative refractive index material would act as a ``perfect lens'' \cite{jbpendry_prl2000, nfang_science2005, xzhang_nmat2008}. This was followed by a variety of related proposals, such as those making use of materials with hyperbolic dispersion \cite{apoddubny_np2013, zliu_science2007, dlu_ncomm2012, sdai_ncomm2015}. Throughout this manuscript, we will refer to any imaging system that uses evanescent waves as an \emph{evanescent microscope}.

The key limitation of evanescent microscopes, of course, is that the evanescent waves quickly decay in the gap between the microscope and the sample. This leads to vanishingly weak signal strengths for all but the shortest focal distances, which we define as the distance between the imaging system and the focal plane. Unfortunately, this runs counter to the desire to have as large of a focal distance as possible for most applications.

There have been surprisingly few attempts to quantify this limitation of evanescent microscopes. The limitations of some specific implementations are well understood, but there are few general results. For example, in the NSOM literature, it is widely understood that a sub-diffraction limited spot produced by a narrow aperture is only maintained for a distance of about one spot width before dissipating \cite{rcdunn_cr1999, yleviatan_jap1986, jmvigoureux_ao1992i, jmvigoureux_ao1992ii, lnovotny_um1998}. Meanwhile, Merlin and Intaraprasonk have proposed schemes which can produce sub-diffraction limited spots at any distance, but exhibit enormous evanescent decay \cite{rmerlin_science2007, vintaraprasonk_ol2009, agrbic_science2008}.

In this work, we derive rigorous bounds on the evanescent decay of fields in the gap between the imaging system and sample. For practical reasons, this space is almost always filled with a conventional medium such as air or water. Evanescent decay across this gap thus only depends upon the transverse spatial frequencies of the waves, and is independent of the method used to create or otherwise amplify these waves. Our analysis thus applies to any focusing or imaging scheme that makes use of evanescent waves, ranging from simple dielectric structures to negative refractive index materials \cite{jbpendry_prl2000, nfang_science2005, xzhang_nmat2008} and hyperbolic metamaterials \cite{apoddubny_np2013, sdai_ncomm2015, zliu_science2007, dlu_ncomm2012}. In particular, we show that any such system will fundamentally be limited to focal distances of a few spot widths, and realistically it will be difficult to image beyond even a single spot width.

The rest of this manuscript is laid out as follows. After discussing the practical implications of our result, we first show that the evanescent decay between the microscope and focal plane depends only upon the spatial frequencies of the fields in the focal plane. We then propose rigorous definitions for the evanescent decay and spot size of arbitrarily shaped beams, since there is no consensus on either in the existing literature. Finally, we derive strict bounds on evanescent decay as a function of spot size in the focal plane.

\begin{figure}[h!tb]
\centering
\includegraphics[width=0.8\columnwidth]{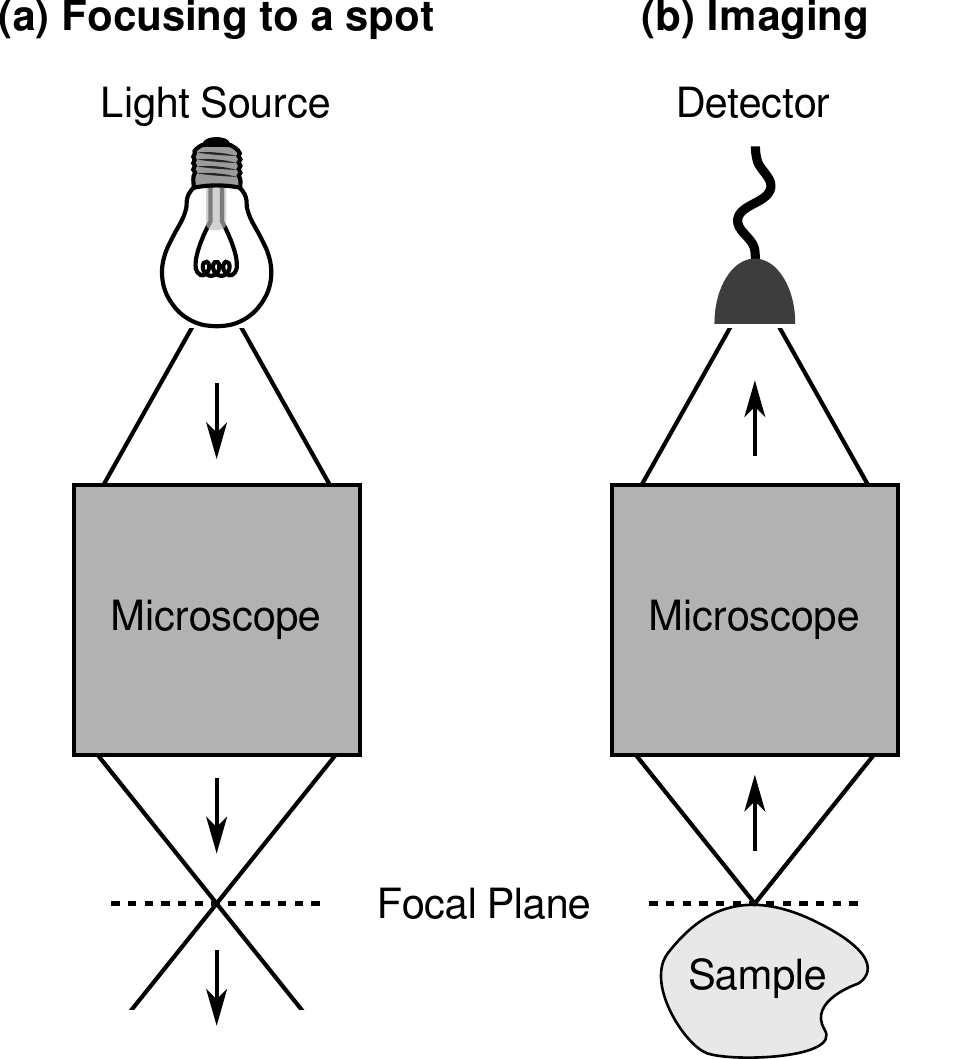}
\caption{A microscope can be used in one of two ways: (a) focusing light to a small spot in the focal plane, and (b) imaging the light emitted or scattered by a sample onto a detector. Due to the reciprocity theorem, these two cases are equivalent to each other.}
\label{fig:sl_focusImagEquiv}
\end{figure}

\emph{Practical implications.} --
An optical microscope can be used in one of two ways. First, as illustrated in figure \ref{fig:sl_focusImagEquiv}a, it can be used to focus light to a small spot in the focal plane. By using evanescent fields, it is possible to produce a spot width $w$ smaller than the diffraction limit of $\sim \lambda / 2$, where $\lambda$ is the wavelength. Unfortunately, as the name implies, these evanescent fields are strongly attenuated as they propagate from the microscope to the focal plane. As we show later in this paper, the electric energy density $\epsilon |\mathbf{E}|^2$ exponentially decays as a function of the distance $L$ from the focusing system:
\begin{align}
\epsilon |\mathbf{E}|^2 \propto \exp\left( - \frac{\pi e}{w} L \right).
\label{eqn:sl_decay_simple}
\end{align}
The decay length $w / \pi e \approx 0.12 w$ is proportional to the spot width $w$. Even if we set the focal distance to be as small as the spot width, equation (\ref{eqn:sl_decay_simple}) implies that the microscope must tolerate three orders of magnitude of decay.

One might optimistically assume that an evanescent microscope could overcome this decay by sufficiently amplifying the evanescent waves. Unfortunately, due to the exponential nature of the decay, the required field strengths at the microscope very quickly exceed the breakdown strengths of any known material. For example, suppose we were to try using an evanescent microscope for lithography. To expose photoresist at the same rate as current lithography tools, we need an electric field of $\sim 10^{2}~\mathrm{V/m}$ at the focal plane.  The required electric field strength at the microscope exceeds the ultimate breakdown strength of all known materials (approximately $10^{10}~\mathrm{V/m}$ \cite{mlenzer_prl1998}) when the focal distance $L$ is only 4 times the spot width $w$.

The second use for a microscope, as illustrated in figure \ref{fig:sl_focusImagEquiv}b, is to image the light emitted or scattered by a sample onto a CCD or other image sensor. It is possible to achieve a resolution better than the diffraction limit of $\sim \lambda / 2$ by detecting the evanescent waves generated by the sample. Unfortunately, as in the previous case, these evanescent waves are quickly attenuated as they propagate from the sample to the microscope, limiting the achievable signal to noise ratio.

To more rigorously quantify the effects of evanescent decay on imaging, we turn to the electromagnetic reciprocity theorem. Broadly speaking, the reciprocity theorem states that a microscope will function identically whether it is used to emit or collect light. As we will see shortly, this implies that equation (\ref{eqn:sl_decay_simple}) not only describes the effects of evanescent decay on focusing light to a point, but also describes its effects when imaging a sample.

For simplicity, let us consider a single detector pixel in the microscope's image sensor. We define the \emph{collection efficiency} of the detector as the power collected by the detector from a point emitter with unit magnitude. Now, suppose we run the microscope in reverse by replacing the detector with a light source. Light will be channeled back through the microscope and focused to a tight spot in the focal plane. Due to the reciprocity theorem, the electric energy density $\epsilon |\mathbf{E}|^2$ of this field is exactly proportional to the collection efficiency of the detector as a function of position (see Supplementary Material \cite{suppl}, section I, for details). We can conclude that the collection efficiency of the microscope drops off exponentially with distance $L$ following equation (\ref{eqn:sl_decay_simple}).

In practice, this implies that evanescent microscopes will have very poor signal to noise ratios when used for imaging. Evanescent decay leads to very poor collection efficiencies and hence weak signal strengths. Although it is hypothetically possible to overcome the weak signal strengths by sufficiently amplifying evanescent waves in the microscope, the exponential nature of evanescent decay implies that the microscope will be vastly more sensitive to objects immediately adjacent to the microscope than at the focal plane. For instance, if the focal distance $L$ is 4 times the spot width $w$ as in our previous example, the microscope will be approximately $10^{15}$ times more sensitive to sources adjacent to the microscope than at the focal plane. This, in turn, makes the microscope susceptible to noise, such as Brownian motion of the water, air, or other medium filling the space between the focal plane and the microscope.

As we have shown, the exceedingly fast attenuation of evanescent waves severely constrains the performance of evanescent microscopes. Even reaching a distance of one spot width requires overcoming 3 orders of magnitude of decay. Indeed, no experiment to date has produced an indisputably sub-diffraction-limited spot further from the focusing system than the spot width \cite{agrbic_science2008}.

\emph{Relationship between evanescent decay and spatial frequencies.} --
We now turn our attention to deriving rigorous bounds on evanescent decay. From this point forwards, we will exclusively study the case of focusing light to a small spot. As discussed earlier, we can use the reciprocity theorem to extend our result to the case of imaging a sample.

\begin{figure}[h!tb]
\centering
\includegraphics[width=\columnwidth]{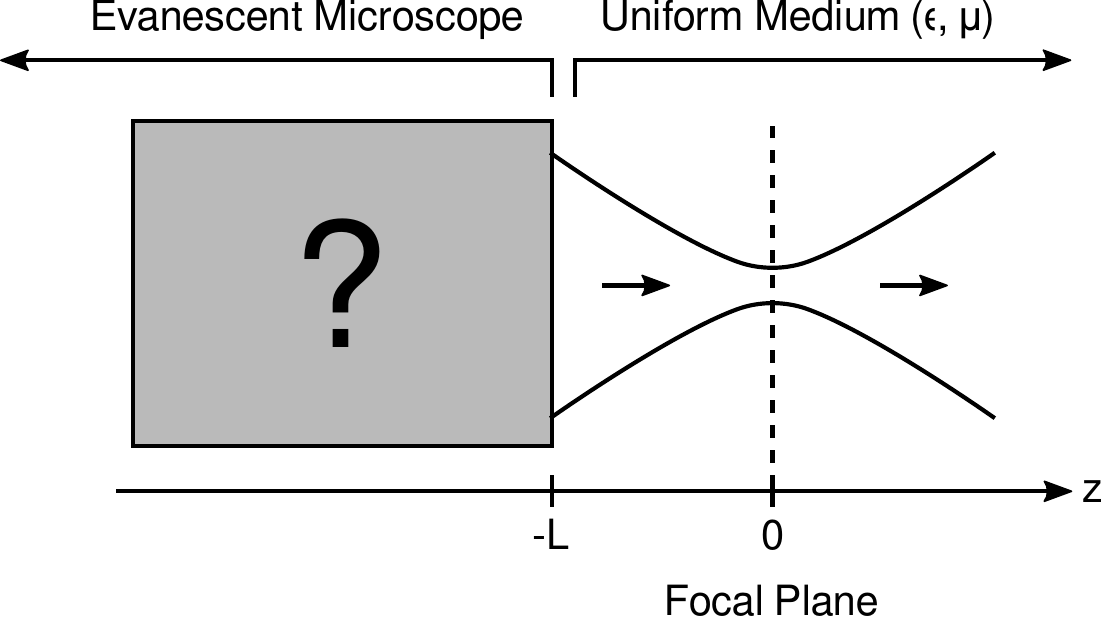}
\caption{Generic focusing problem. Light is emitted by an evanescent microscope  and is focused to a small spot in the focal plane $z = 0$. The microscope is separated from the focal plane by focal distace $L$. We assume that the half space $z > -L$ outside of the microscope is filled with a uniform, lossless medium described by a positive and scalar permittivity $\epsilon$ and permeability $\mu$. In addition, we assume that there are no propagating or evanescent waves incident from $z \rightarrow +\infty$.}
\label{fig:sl_focusProb} 
\end{figure}

The specific system we will consider is illustrated in figure \ref{fig:sl_focusProb}. Light is emitted by an evanescent microscope in the $+z$ direction, and focused down to a small spot in the focal plane at $z = 0$. The microscope and the focal plane are separated by distance $L$, which we define as the \emph{focal distance}. The entire microscope is confined to the half space $z < -L$.

We will only concern ourselves with the behaviour  of fields in the half space $z > -L$ outside of the microscope. Our key assumption is that entire volume $z > -L$ is filled with a uniform, isotropic, and lossless medium, which represents a conventional medium such as air or water. In this region, each vectorial component of the $\mathbf{E}$ and $\mathbf{H}$ fields satisfies the scalar Helmholtz equation \cite{rfharrington_2001_helmholtz}
\begin{align}
\nabla^2 f + k_0^2 f = 0.
\label{eqn:sl_Helmholtz}
\end{align}
Here, $f(x,y,z)$ is a scalar field representing any vector component of the $\mathbf{E}$ and $\mathbf{H}$ fields, and $\nabla^2 = \partial_x^2 + \partial_y^2 + \partial_z^2$ is the Laplacian. The wavenumber $k_0$ is related to the wavelength $\lambda$ of light in the medium by $k_0 = 2 \pi / \lambda$. 

The fields in the entire half space $z > -L$ outside of the microscope are in fact uniquely determined by the fields in the focal plane $z = 0$. This includes the space $-L < z < 0$ between the microscope and the focal plane. To show this, we first decompose the field $f(x,y,z)$ into its constituent plane waves by taking a spatial Fourier transform over the focal plane:
\begin{align}
\hat{f}(k_x, k_y) = \frac{1}{2 \pi} \iint f(x,y,0) e^{-i (k_x x + k_y y)} dx dy.
\label{eqn:sl_ft}
\end{align}
The quantity $\hat{f}(k_x, k_y)$ represents the amplitude of a plane wave of the form
\begin{align}
f_{\mathbf{k}}(x,y,z) = \hat{f}(k_x, k_y) e^{i (k_x x + k_y y + k_z z)}
\label{eqn:sl_planewave}
\end{align}
with purely real transverse spatial frequencies $k_x$ and $k_y$. Here, we have assumed the fields have a time dependence of $e^{-i \omega t}$.

The longitudinal wavenumber $k_z$ can be found by substituting (\ref{eqn:sl_planewave}) into the Helmholtz equation (\ref{eqn:sl_Helmholtz}). This yields
\begin{align}
k_z = \sqrt{\frac{4 \pi^2}{\lambda^2} - k_x^2 - k_y^2},
\label{eqn:sl_kz}
\end{align}
where we have ensured that the waves propagate in the $+z$ direction by setting $\real(k_z) > 0$ and $\imag(k_z) > 0$. Tranverse spatial frequencies within the circle $k_x^2 + k_y^2 < 4 \pi^2/\lambda^2 $ correspond to propagating waves with real $k_z$. Conventional microscopes, which only make use of propagating waves, are thus limited to relatively small spatial frequencies in the focal plane. Meanwhile, spatial frequencies outside this circle with $k_x^2 + k_y^2 > 4 \pi^2/\lambda^2 $ correspond to evanescent waves with imaginary $k_z$. It is these evanescent waves that evanescent microscopes seek to exploit to achieve high spatial resolution. Unfortunately, evanescent waves decay quite quickly in the $z$ direction, with a decay length $z_D$ of
\begin{align}
z_D = \frac{1}{\imag(k_z)} = \frac{1}{\sqrt{ k_x^2 + k_y^2 - \frac{4 \pi^2}{\lambda^2}}}.
\label{eqn:sl_decaylength}
\end{align}
In the limit of high spatial frequencies, the decay length $z_D$ is inversely proportional to the transverse spatial frequency $\sqrt{k_x^2 + k_y^2}$. 

To reconstruct the field at any point $(x, y, z)$ in the half space $z > -L$ outside the microscope, we simply propagate the plane waves to the given point and take the inverse spatial Fourier transform:
\begin{align}
f(x, y, z) =  \frac{1}{2 \pi} \iint \hat{f}(k_x,k_y) e^{i (k_x x + k_y y + k_z z)} dk_x dk_y.
\label{eqn:sl_ift}
\end{align}
This reconstruction is unique, as we show in the Supplementary Information \cite{suppl}, section II. In addition, we can see from equation (\ref{eqn:sl_decaylength}) that the decay length of an evanescent wave depends only upon its transverse spatial frequencies. Thus, evanescent decay between a microscope and its focal plane is uniquely determined by the transverse spatial frequencies present in the focal plane. Furthermore, we know from the uncertainty principle that smaller focal spots contain higher spatial frequencies, which in turn have faster evanescent decay. We can therefore conclude that there is a fundamental tradeoff between spot size and evanescent decay. We will spend the rest of this manuscript deriving rigorous bounds on this tradeoff.

\begin{figure}[h!tb]
\centering
\includegraphics[width=\columnwidth]{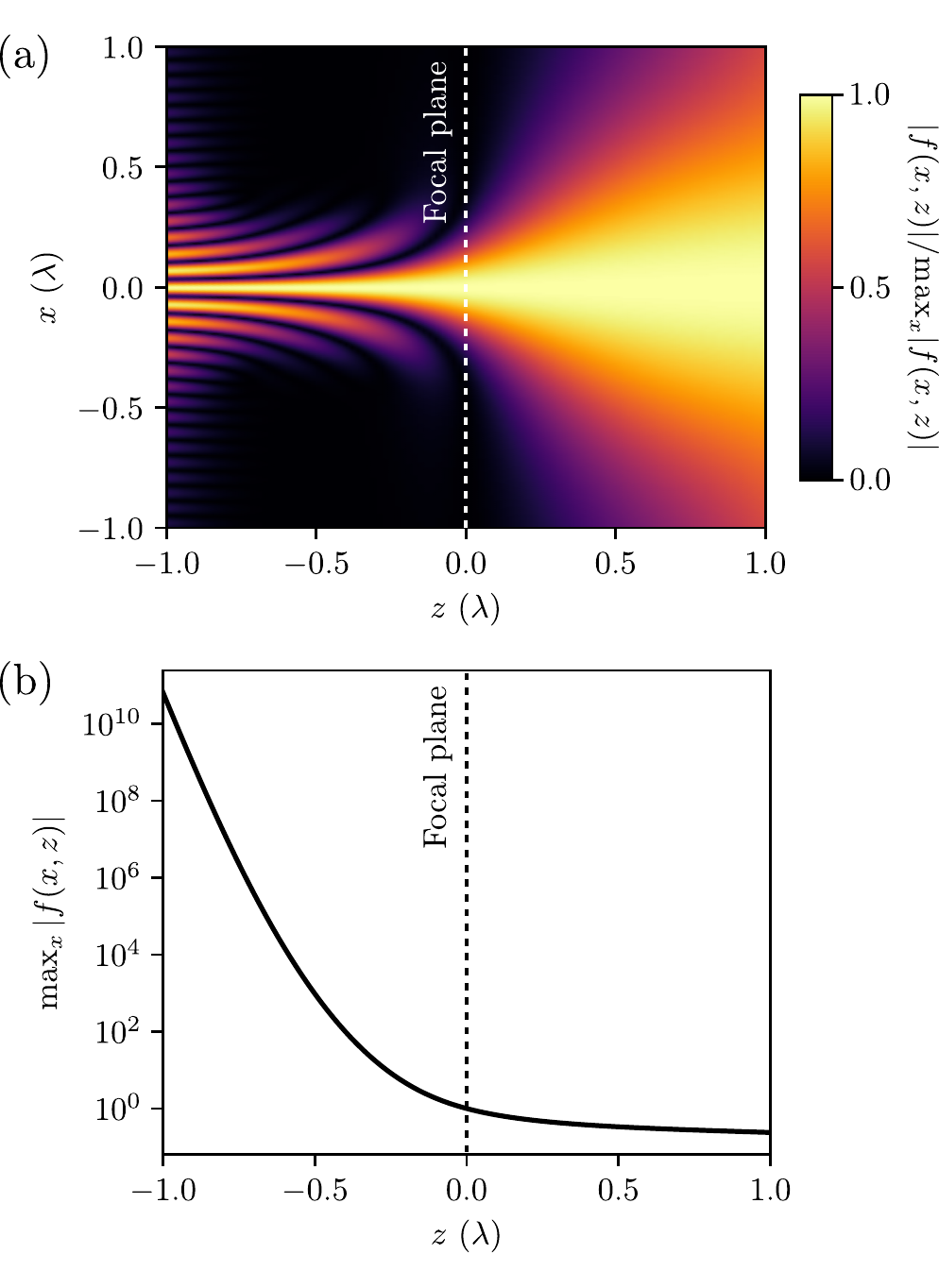}
\caption{Reconstructed fields in two dimensions for a $0.4~\lambda$ diameter Gaussian focal spot. The focal plane is located in the $z = 0$ plane, indicated by the dashed line, and the waves propagate from left to right in the $+z$ direction. We have plotted (a) the normalized field magnitude $|f(x,z)| / \max_x |f(x,z)|$, and (b) the peak electric field $\max_x |f(x,z)|$ as a function of $z$.  We can see that the field decays extremely rapidly between the microscope and focal plane ($z < 0$) due to the presence of evanescent waves. On the far side of the focal plane ($z > 0$), the fields are dominated by propagating waves which do not decay.}
\label{fig:sl_spot_extrap}
\end{figure}

\emph{Gaussian beam example} --
The evanescent decay of fields between the microscope and focal plane is extremely fast, as illustrated in the following example. For simplicity, we will restrict ourselves to the case of a two dimensional beam, i.e. one where the fields are a function of only $x$ and $z$. Suppose this beam has as a Gaussian profile in the focal plane $z = 0$,
\begin{align}
f(x, 0) = \exp\left(- \frac{x^2}{\sigma^2} \right),
\end{align}
with a spot diameter $2 \sigma$ of $0.4 \lambda$, which is only slightly smaller than the diffraction limit of $\sim \lambda / 2$. Given the fields in the focal plane $z = 0$, we can reconstruct the fields everywhere outside of the microscope using equations (\ref{eqn:sl_ft}), (\ref{eqn:sl_kz}), and (\ref{eqn:sl_ift}). The resulting fields are plotted in figure \ref{fig:sl_spot_extrap}. Despite the relatively large size of the focal spot, the field strength drops by more than 10 orders of magnitude in the span of a single wavelength.

\emph{Quantifying Evanescent Decay.} --
Our next step is to quantify the evanescent decay of arbitrary field distributions. To this end, we define the \emph{field energy} $U(z)$ in any plane parallel to the focal plane as
\begin{align}
U(z) = \iint \left| f(x, y, z) \right|^2 \, dx \, dy.
\end{align}
We then define the \emph{field energy decay} $D$ as the ratio of the field energy at the microscope $z = -L$ and the focal plane $z = 0$:
\begin{align}
D = \frac{U(0)}{U(-L)}.
\label{eqn:sl_D_def}
\end{align}
A field consisting only of propagating waves has $D = 1$. However, $0 < D < 1$ when evanescent waves are present.

For convenience, we will assume that the field $f$ is normalized such that the focal plane field energy
\begin{align}
U(0) = \iint \left| f(x, y, 0) \right|^2 \, dx \,dy = 1
\end{align}
for the rest of this manuscript. This allows us to simplify our expression for field energy decay to
\begin{align}
\frac{1}{D} = \iint \left| f(x, y, -L) \right|^2 \, dx \, dy.
\label{eqn:sl_D_1}
\end{align}
Using Plancherel's theorem and (\ref{eqn:sl_ft}), this can be rewritten in terms of the spatial frequency spectrum of the focal plane fields $\hat{f}(k_x, k_y)$ as
\begin{align}
\frac{1}{D} = \iint \left| \hat{f}(k_x, k_y) \right|^2 \, \left| e^{- i k_z L} \right|^2 \, dk_x \, dk_y.
\label{eqn:sl_D_simple}
\end{align}
We can see that the change in energy for any given transverse spatial frequency is
\begin{align}
&\left| e^{- i k_z L} \right|^2 \nonumber\\
&=
\begin{cases} 
1 & \mathrm{for} \; k_x^2 + k_y^2 \leq  \frac{4 \pi^2}{\lambda^2} \\
\exp \left( 2 L \sqrt{k_x^2 + k_y^2 - \frac{ 4 \pi^2}{\lambda^2}} \right) & \mathrm{otherwise.}
\end{cases}
\label{eqn:sl_u_def}
\end{align}
For propagating waves within the circle $ k_x^2 + k_y^2 \leq  4 \pi^2 / \lambda^2$, the energy stays constant. However, for evanescent waves outside this circle, the change in energy is an exponential function of the focal distance $L$ and transverse spatial frequencies $k_x$ and $k_y$.

\emph{Definition of Spot Size.} --
We now turn our attention to finding a robust definition for ``spot size'' in the focal plane. Our goal is to find a definition for spot size that applies to arbitrary beam profiles, and remains well behaved for any commonly encountered beam profiles.

At this point in our discussion, it is useful to draw a distinction between two-dimensional (2D) and three-dimensional (3D) focusing. In 2D focusing, the fields $f(x,z)$ depend only upon $x$ and $z$, and are characterized by a spot width $w$ in the focal plane $z = 0$. Meanwhile, in 3D focusing, the fields $f(x, y, z)$ depend on all three spatial variables and are characterized by a spot area $A$.

Unfortunately, commonly used measures of spot width and area in optics are not particularly robust. The full-width half-maximum (FWHM) is only useful for well behaved spots such as Gaussians. The other commonly used measure, the root mean square (RMS) spot width, is more generally applicable to arbitrary beams. In the case of 2D focusing, the RMS spot width is defined in terms of the focal plane fields $f(x, 0)$ as
\begin{align}
w_{RMS} = \int \left| f(x,0) \right|^2 x^2 dx - \left( \int \left| f(x,0) \right|^2 x \, dx \right)^2.
\end{align}
Unfortunately, the RMS spot width diverges to infinity for any aperture diffracted beam such as the ubiquitous Airy disk \cite{maporras_oc1994}.

A more robust definition of spot size, which remains well defined for heavy-tailed distributions such as the Airy disk, comes from the information theory concept of differential or continuous entropy \cite{ahyvarinen_1997, cmgoldie_1991, avlazo_ieeetoit1978}. This metric has has previously been proposed as a measure of laser beam spot size \cite{maporras_ao1995}, and is widely used as a measure of uncertainty in quantum mechanics \cite{ibialynickibirula_2011, swehner_njp2010, jpcoles_rmp2017}. In the 2D focusing case, we first define the differential entropy as
\begin{align}
H_w = - \int \left| f(x, 0) \right|^2 \ln \left| f(x, 0) \right|^2 dx
\label{eqn:sl_def_Hw}
\end{align}
and use it to define the entropic spot width $w$:
\begin{align}
w = \exp\left(H_w\right).
\label{eqn:sl_def_w}
\end{align}
Similarly, we can define a spot width $\hat{w}$ in k-space as
\begin{align}
\hat{w} &= \exp\left( - \int \left| \hat{f}(k_x) \right|^2 \ln \left| \hat{f}(k_x) \right|^2 dk_x
\right).
\label{eqn:sl_def_wk}
\end{align}
The spot widths in real space and k-space spot are related by the uncertainty principle \cite{ibialyniki-birula_cmp1975}
\begin{align}
w \, \hat{w} \geq e^{(1 + \ln \pi)}.
\label{eqn:sl_uncertainty_w}
\end{align}

The extension to 3D focusing, where we have a spot area rather than spot width, is trivial: we simply integrate over two dimensions rather than one. The spot areas in real space $A$ and k-space $\hat{A}$ are
\begin{align}
A &= \exp\left( - \int \left| f(x, y, 0) \right|^2 \ln \left| f(x, y, 0) \right|^2 dx \, dy\right)
\label{eqn:sl_def_A} \\
\hat{A} &= \exp\left( - \int \left| \hat{f}(k_x, k_y) \right|^2 \ln \left| \hat{f}(k_x, k_y) \right|^2 dk_x \, dk_y \right),
\label{eqn:sl_def_Ak}
\end{align}
and the uncertainty principle becomes
\begin{align}
A \, \hat{A} \geq e^{2(1 + \ln \pi)}.
\label{eqn:sl_uncertainty_A}
\end{align}

The entropic spot size can be understood as the ``width'' or ``area'' filled by the distribution $|f(x,0)|^2$ or $|f(x, y, 0)|^2$  \cite{mjwhall_pra1999}. For example, if the field in the focal plane is a rectangular function
\begin{align}
f(x, 0) = \frac{1}{\sqrt{L}} 
\begin{cases}
1, & 0 < x < L\\
0, & \text{otherwise,}
\end{cases}
\end{align}
then the real-space spot width $w$ is exactly $L$. The entropic spot size also matches well with standard definitions of spot size for commonly encountered beam profiles. If we consider the Gaussian profile
\begin{align}
f(x, 0) = \frac{1}{\sqrt[4]{\pi \sigma^2}} \exp\left( - \frac{x^2}{\sigma^2} \right),
\end{align}
then the spot width $w =  \sigma \sqrt{\pi e / 2} \approx 2.07\sigma$, which is close to the standard definition of $2\sigma$ for a Gaussian beam. A more challenging example is the one-dimensional analogue of the Airy disk, the normalized sinc function
\begin{align}
f(x, 0) = \frac{\sin(\pi x)}{\pi x}.
\end{align}
The sinc function's RMS spot width diverges to infinity, but has a well defined entropic spot width $w \approx 2.33$. The latter value matches our intuitive expectations since the first zeros of the sinc function are located at $x = \pm 1$.

\emph{Bounds on evanescent decay.} --
Now that we have precise defintions for both evanescent decay and spot size, we are finally in a position to derive rigorous bounds on evanescent decay. For clarity, we will focus on the case of 2D focusing in the main text; the 3D case is a straightforward extension and is discussed in the supplementary information.

We would like to find an upper bound on the field energy decay $D$ from the microscope to the focal plane, which will be a function of both the focal distance $L$ and spot width $w$ in the focal plane. To accomplish this, we first find an upper bound on the k-space spot width $\hat{w}$ for given values of $D$ and $L$ by solving the convex optimization problem
\begin{alignat}{3}
&\maximize  \quad & \ln \hat{w} &= -\textstyle\int \rho(k_x) \ln \rho(k_x) \; dk_x  \nonumber \\
&\subjectto \quad & 0 &\leq \rho(k_x)  \nonumber \\
&                 & 1 &= \textstyle\int \rho(k_x) \;dk_x \nonumber \\
&                 & \frac{1}{D} &= \textstyle\int \rho(k_x) \; \left| e^{- i k_z L} \right|^2 \; dk_x .
\label{eqn:sl_opt_prob}
\end{alignat}
Here, we have defined $\rho(k_x) = \left|\hat{f}(k_x)\right|^2$, and made use of our definitions of field energy decay (\ref{eqn:sl_D_simple}) and spot width (\ref{eqn:sl_def_Hw}), (\ref{eqn:sl_def_w}). In the Supplementary Material \cite{suppl}, section III, we show that the optimal distribution $\rho^\ast(k_x)$ that solves (\ref{eqn:sl_opt_prob}) is
\begin{align}
\rho^\ast(k_x) = N \exp\left(- \beta \left| e^{- i k_z L} \right|^2 \right),
\label{eqn:sl_p_opt}
\end{align}
where $N$ and $\beta$ are positive constants that can be solved for numerically using the constraints in the original problem (\ref{eqn:sl_opt_prob}). Next, as we discuss in the Supplementary Material \cite{suppl}, section IV, we can use the uncertainty principle (\ref{eqn:sl_uncertainty_w}) to map the upper bound on k-space spot width $\hat{w}$ to a lower bound on real-space spot width $w$. Finally, since the lower bound on $w$ is an increasing function of $D$, we can invert the relationship to obtain an upper bound on $D$ as a function of $w$.

\begin{figure}[h!tb]
\centering
\includegraphics[width=\columnwidth]{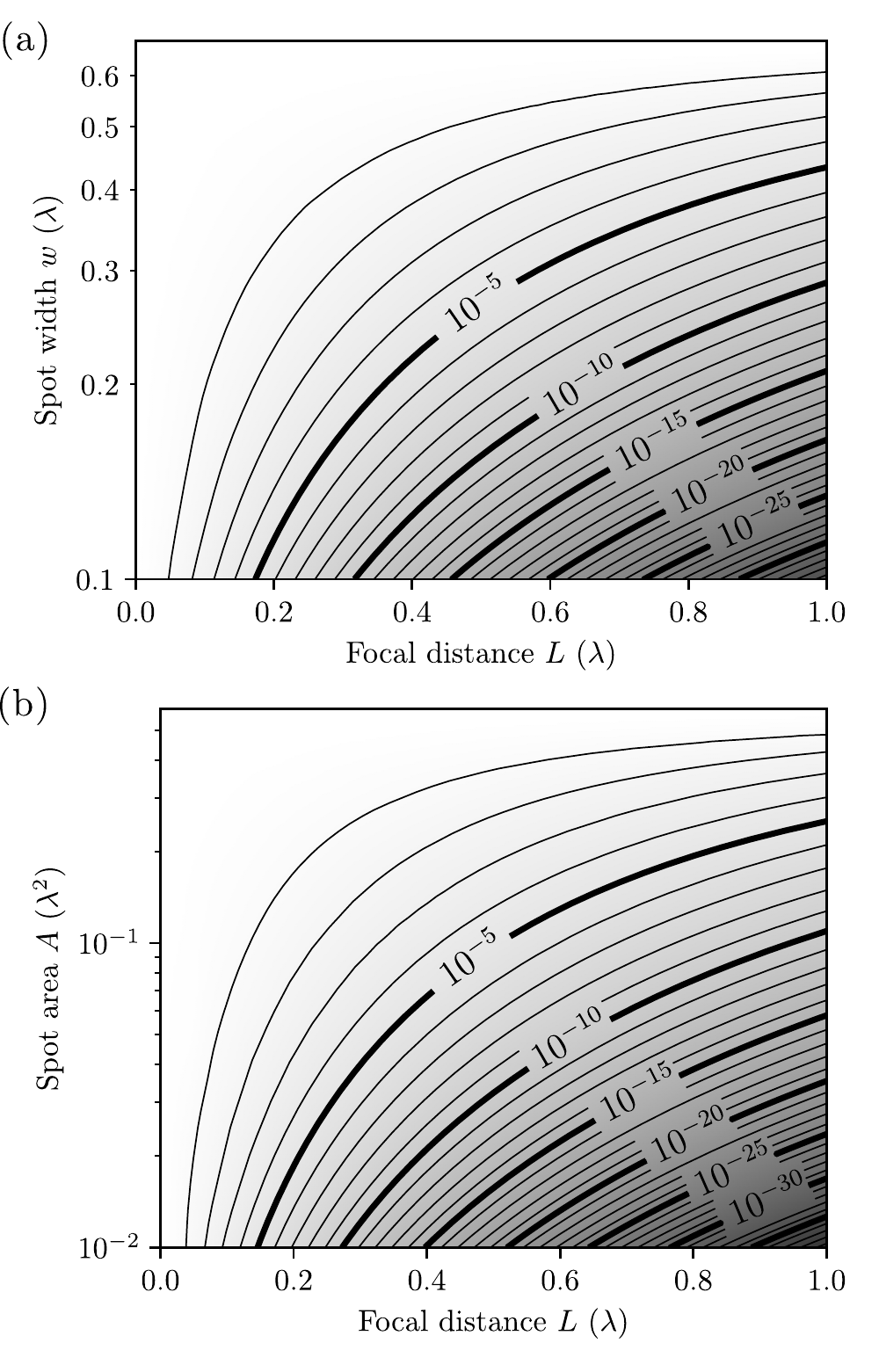}
\caption{Best-case field energy decay $D$ as a function of focal distance $L$ and spot size, for (a) two-dimensional and (b) three-dimensional focusing. The fine and coarse contours indicate steps of $10^{-1}$ and $10^{-5}$ respectively. The upper limit of spot size in these plots is equal to the smallest possible spot size using only propagating fields. In the 2D case, this corresponds to a width of $e \lambda / 4 \approx 0.680 \, \lambda$, whereas in the 3D case, this corresponds to an area of $e^2 \lambda^2 / 4 \pi \approx 0.588 \, \lambda^2$.} 
\label{fig:sl_decay}
\end{figure}

In figure \ref{fig:sl_decay}, we have plotted the best-case field-energy decay $D$ as a function of the real-space spot width $w$ and focal distance $L$. Even for spots which are only modestly smaller than the diffraction limit, the field energy $D$ decays extremely quickly as the focal distance increases.

\emph{Far-field regime.} --
In the far-field regime where conventional optical microscopes operate, the fields in the focal plane can only consist of propagating waves. This corresponds to limiting the transverse wavevectors to the circle
\begin{align}
k_x^2 + k_y^2 \leq \left(\frac{2 \pi}{\lambda} \right)^2.
\label{eqn:sl_farfield_circle}
\end{align}
As we show in the Supplementary Material, section V \cite{suppl}, the real space spot width is then limited to
\begin{align}
w \geq \frac{e}{4} \lambda \approx 0.680 \, \lambda.
\end{align}
This nicely corresponds to the common rule of thumb that the resolution of a conventional microscope is limited to roughly half a wavelength. Similarly, the real-space spot area $A$ must be greater than
\begin{align}
A \geq \frac{e^2}{4 \pi} \lambda^2 \approx 0.588 \, \lambda^2.
\end{align}

\emph{Deep subwavelength regime.} --
For spots that are even modestly smaller than the diffraction limit, the transverse wavevectors extend well past the circle $k_x^2 + k_y^2 < \left(2 \pi / \lambda \right)^2$. In addition, from our expressions for field energy decay (\ref{eqn:sl_D_simple}) and (\ref{eqn:sl_u_def}), it is clear that the contributions from the largest transverse wavevectors dominate the total evanescent decay. We can thus make the approximation
\begin{align}
\left| e^{- i k_z L} \right|^2  \approx \exp\left( 2 L \sqrt{k_x^2 + k_y^2} \right),
\end{align}
allowing us to derive simplified analytic bounds on the field energy decay $D$. We present the key results here; the detailed derivations are located in the Supplementary Material, section VI \cite{suppl}.

In 2D focusing, we obtain the approximate upper bound
\begin{align}
D \lessapprox \left(\frac{L}{\delta_{2D}} + 1 \right) \exp\left( - \frac{L}{\delta_{2D}} + 1 - \gamma \right),
\label{eqn:sl_constr_2d_approx}
\end{align}
where the decay length $\delta_{2D}$ is proportional to the spot width $w$:
\begin{align}
\delta_{2D} = \frac{w}{\pi e} \approx 0.12 \, w.
\label{eqn:sl_constr_2d_R}
\end{align}
Here, $\gamma \approx 0.577$ is the Euler-Mascheroni constant. This approximation is valid when the focal distance $L$ is sufficiently large such that
\begin{align}
L \gg \frac{2 - \gamma}{e \pi} \, w \approx 0.17 \, w.
\label{eqn:sl_small_beta_2d}
\end{align}
Thus, for sufficiently large focal distances, the fields decay roughly exponentially from the microscope to the focal plane, with a decay length $\delta_{2D} \approx 0.12 \, w$.

In the 3D case, the upper bound on $D$ is approximately 
\begin{align}
D \lessapprox \frac{L}{2 \, \delta_{3D}} \exp\left(- \frac{L}{\delta_{3D}} - \gamma \right),
\label{eqn:sl_constr_3d_approx}
\end{align}
where the decay length $\delta_{3D}$ is roughly proportional to the square root of the spot area $A$:
\begin{align}
\delta_{3D} &= L \left(-1 + \sqrt{1 + \frac{4 \pi e^2 L^2}{A}} \right)^{-1} \nonumber\\
&\approx \frac{\sqrt{A}}{2 \sqrt{\pi} e} \approx 0.10 \sqrt{A}.
\label{eqn:sl_constr_3d_R}
\end{align}
This approximation is again valid when the focusing distance $L$ is sufficiently large, with the slightly modified constraint
\begin{align}
L \gg \frac{\sqrt{A}}{2 e \sqrt{\pi}} \approx 0.10 \sqrt{A} .
\label{eqn:sl_small_beta_3d}
\end{align}
Thus, in three-dimensional focusing, the fields also decay exponentially from the microscope to the focal plane, with a decay length $\delta_{3D} \approx 0.10 \sqrt{A}$. 

\emph{Conclusion.} --
We have found strong constraints on any imaging or focusing scheme that relies on evanescent waves. In particular, we have derived rigorous bounds on the decay of electromagnetic energy from the evanescent microscope to the focal plane. In two-dimensional focusing, the decay length is only $w / \pi e \approx 0.12 w$, and is proportional to the spot width $w$. In three-dimensional focusing, the decay length is $\sqrt{A} / 2 e \sqrt{\pi} \approx 0.10 \sqrt{A}$, scaling as the square root of the spot area $A$. Our results strongly constrain what can be achieved by any evanescent imaging scheme.

\medskip

We would like to thank Shanhui Fan, David A. B. Miller, and Rahul Trivedi for their helpful comments. In addition, we are grateful to the Gordon and Betty Moore Foundation for supporting this work.

%merlin.mbs apsrev4-1.bst 2010-07-25 4.21a (PWD, AO, DPC) hacked
%Control: key (0)
%Control: author (8) initials jnrlst
%Control: editor formatted (1) identically to author
%Control: production of article title (-1) disabled
%Control: page (0) single
%Control: year (1) truncated
%Control: production of eprint (0) enabled
%

%%%%%%%%%% Merge with supplemental materials %%%%%%%%%%
\clearpage
\onecolumngrid

\begin{center}
\textbf{\large Supplemental Information}
\end{center}

\twocolumngrid
%%%%%%%%%% Merge with supplemental materials %%%%%%%%%%
%%%%%%%%%% Prefix a "S" to all equations, figures, tables and reset the counter %%%%%%%%%%
\setcounter{equation}{0}
\setcounter{figure}{0}
\setcounter{table}{0}
\setcounter{page}{1}
\makeatletter
\renewcommand{\theequation}{S\arabic{equation}}
\renewcommand{\thefigure}{S\arabic{figure}}
\renewcommand{\bibnumfmt}[1]{[S#1]}
\renewcommand{\citenumfont}[1]{S#1}
%%%%%%%%%% Prefix a "S" to all equations, figures, tables and reset the counter %%%%%%%%%%

\section{Equivalence of Imaging and Focusing}
\label{sec:sl_imagfocusequiv}
As we discussed in the main text, a microscope can either be used to focus light to a small spot, or image the light from a sample onto an image sensor. Here, we will rigorously show that these cases are equivalent to each other due to the reciprocity theorem.

Our first step is to precisely define the imaging configuration. Suppose we are interested in imaging a current density distribution $\mathbf{J}_i(\mathbf{r})$, which produces electric fields $\mathbf{E}_i(\mathbf{r})$. This is essentially equivalent to imaging the light scattered by a sample, since we can replace the far field light source with a current distribution in the focal plane using the equivalence principle \cite{rfharrington_2001}. Our microscope is some arbitrary distribution of permittivity $\epsilon(\mathbf{r})$ and permeability $\mu(\mathbf{r})$, which acts to focus the light on an image sensor.

Next, consider a single photodetector in the image sensor array. For now, we assume that this is a coherent single-mode detector, by which we mean the signal amplitude $A$ is given by the overlap integral
\begin{align}
A = \iiint \mathbf{S}(\mathbf{r}) \cdot \mathbf{E}_i(\mathbf{r}) \, d\mathbf{r},
\label{eqn:sl_det_response}
\end{align}
where $\mathbf{S}(\mathbf{r})$ is the detector sensitivity. We can create a detector of this form by, for example, placing one end of a single-mode optical fiber in the back focal plane of the microscope, and attaching a coherent optical receiver to the other end of the fiber.

We construct an equivalent focusing configuration by creating a new system that is identical in all respects to the imaging problem, except we choose the current density $\mathbf{J}_f(\mathbf{r})$ to be equal to the detector sensitivity $\mathbf{S}(\mathbf{r})$. This will propagate light backwards through our microscope, focusing light to a small spot in the focal plane. We denote the electric field produced by $\mathbf{J}_f(\mathbf{r})$ as $\mathbf{E}_f(\mathbf{r})$.

Since the imaging and focusing configurations are described by the same $\epsilon$ and $\mu$ distributions, we can apply the Lorentz reciprocity theorem \cite{rfharrington_2001} to yield
\begin{align}
\iiint \mathbf{E}_i(\mathbf{r}) \cdot \mathbf{J}_f(\mathbf{r}) \, d\mathbf{r} = \iiint \mathbf{E}_f(\mathbf{r}) \cdot \mathbf{J}_i(\mathbf{r}) \, d\mathbf{r},
\label{eqn:sl_reciprocity}
\end{align}
where we have integrated over all space. However, the left hand side of (\ref{eqn:sl_reciprocity}) is equal to our detector signal amplitude $A$ since $\mathbf{J}_f(\mathbf{r}) = \mathbf{S}(\mathbf{r})$, and we obtain
\begin{align}
A = \iiint \mathbf{E}_f(\mathbf{r}) \cdot \mathbf{J}_i(\mathbf{r}) \, d\mathbf{r}.
\end{align}
Thus, our detected signal $A$ is given by the overlap of the current source $\mathbf{J}_i(\mathbf{r})$ we are trying to image, and the electric field $\mathbf{E}_f(\mathbf{r})$ produced by using the detector sensitivity as a source.  Equivalently, if we have a point emitter at a given position $\mathbf{r}$, the signal power $|A|^2$ will be proportional to $|\mathbf{E}_f(\mathbf{r})|^2$, the square of the local electric field  in the focusing configuration.

We can handle systems which incorporate non-reciprocal materials by using a more general form of the reciprocity theorem \cite{recollin_1991}. When the permittivity $\epsilon(\mathbf{r})$ or permeability $\mu(\mathbf{r})$ tensors are asymmetric, the $\epsilon(\mathbf{r})$ and $\mu(\mathbf{r})$ tensors in the focusing configuration are the transpose of those in the imaging configuration.

Finally, we can extend our result to incoherent detectors, such as bare photodiodes and CCD pixels, by expanding the incoherent detector's signal power $P$ in terms of the signal amplitude of a set of coherent detectors:
\begin{align}
P = \sum_a \left| A^{(a)} \right|^2 
= \sum_a \left| \iiint \mathbf{E}_i(\mathbf{r}) \cdot \mathbf{S}^{(a)}(\mathbf{r}) \, d\mathbf{r} \right|^2.
\end{align}
The imaging resolution of an incoherent detector is strictly worse than a coherent detector since its response is the incoherent sum of a number of coherent detectors. 

\section{Uniqueness of Reconstruction}
Our paper heavily relies on the assumption that the fields on any plane $z = C$ for positive $C$ uniquely determine the fields on the entire half space $z > 0$, given appropriate conditions. Here, we will prove that this assumption is true. 

Consider a scalar field $f : \mathbb{R}^3 \rightarrow \mathbb{C}$ that satisfies the Helmholtz equation
\begin{align}
\nabla^2 f(x,y,z) + k_0^2 f(x,y,z) = 0
\label{eqn:sl_Helmholtz_suppl}
\end{align}
in the half space $z > 0$. The wavenumber $k_0$ is any complex scalar such that $\imag(k_0) > 0$, which corresponds to wave propagation in a uniform, isotropic, and dissipative medium. Now, suppose we are given Dirichlet data on the $z=0$ plane:
\begin{align}
f(x,y,0) = g(x,y).
\label{eqn:sl_dirichlet_z=0}
\end{align}
We assume that the boundary data $g$ is in the Sobolev space $H^{1/2}$, which is defined as \cite{mlassas_ip1998}
\begin{align}
H^s = \left\{ g: \iint \left| \hat{g}(k_x, k_y) \right|^2 \left(1 + k_x^2 + k_y^2\right)^s dk_x \, dk_y < \infty \right\}.
\label{eqn:sl_def_sobelev}
\end{align}
Here $\hat{g}(k_x, k_y)$ denotes the Fourier transform of $g(x,y)$. We further assume that we have radiation boundary conditions on the half space $z > 0$, which is equivalent to the less precise statement that there are no fields incident from $z \rightarrow +\infty$.

This half-space wave propagation problem is known to have a unique solution \cite{mcheney_ip1995, mlassas_ip1998}. We explicitly constructed the solution to this problem in the main text, which we found to be
\begin{align}
f(x,y,z) = \iint \hat{g}(k_x, k_y) e^{i(k_x x + k_y y + k_z z )} dk_x \, dk_y.
\label{eqn:sl_ift_suppl}
\end{align}
Here, the longitudinal wavenumber $k_z$ is given by
\begin{align}
k_z = \sqrt{k_0^2 - k_x^2 - k_y^2}.
\label{eqn:sl_kz_suppl}
\end{align}
Since $g(x,y)$ is in the Sobolev space $H^{1/2}$, equation (\ref{eqn:sl_ift_suppl}) is always well behaved.

Now, suppose we are instead given Dirichlet data on some other plane $z = C$ with positive $C$. We can explicitly write this condition as
\begin{align}
f(x,y,C) = h(x,y).
\label{eqn:sl_dirichlet_z=L}
\end{align}
The fields on the $z = 0$ plane, $g(x,y)$, are related to the fields on the $z = C$ plane, $h(x,y)$ by equation (\ref{eqn:sl_ift_suppl}), yielding
\begin{align}
h(x,y) = \iint \hat{g}(k_x, k_y) e^{i(k_x x + k_y y + k_z C )} dk_x \, dk_y.
\label{eqn:sl_ift_suppl_z=C}
\end{align}
However, since the Fourier transform is invertible, $\hat{g}(k_x, k_y)$ and $g(x,y)$ are uniquely determined by $h(x,y)$. Due to our earlier uniqueness theorem, $f$ is thus uniquely determined on the entire half space $z > 0$. Finally, to obtain the lossless case studied in the main text, we simply take the limit $\imag(k_0) \rightarrow 0$.

\section{Entropy Maximization Problem}
\label{sec:sl_entropyopt}
The Shannon entropy of a continuous probability distribution $\rho:\mathbb{R}^n \rightarrow \mathbb{R}$ is defined as \cite{cmgoldie_1991}
\begin{align}
H(\rho) = - \int \rho(\mathbf{x}) \ln \rho(\mathbf{x}) \, d\mathbf{x}.
\end{align}
The constrained entropy minimization problem
\begin{alignat*}{3}
&\maximize  \quad && H(\rho) \\
&\subjectto \quad && 0 \leq \rho(\mathbf{x})  \\
&                 && 1 = \textstyle\int \rho(\mathbf{x}) \, d\mathbf{x}\\
&                 && U = \textstyle\int \rho(\mathbf{x}) \, u(\mathbf{x}) \, d\mathbf{x}
\end{alignat*}
is solved by the distribution $ \rho^\ast(\mathbf{x}) = \exp\left(- \alpha - \beta u(\mathbf{x}) \right) $ if there exist real $\alpha$ and $\beta$ that satisfy these constraints.

\begin{proof}
Suppose we have any distribution $\rho$ that satisfies the constraints. Its entropy is
\begin{align*} 
H(\rho) = - \int \rho \ln \left( \rho \, \frac{\rho^\ast}{\rho^\ast} \right) \, d\mathbf{x} 
= - \int \rho \ln \rho^\ast \, d\mathbf{x} - \int \rho \ln \frac{\rho}{\rho^\ast} \, d\mathbf{x} .
\end{align*}
Applying Gibb's inequality \cite{cmgoldie_1991} 
\begin{align*}
\int \rho \ln \frac{\rho}{\rho^\ast} \, d\mathbf{x} \geq 0
\end{align*}
yields the upper bound
\begin{align*}
H(\rho) \leq - \int \rho \ln \rho^\ast \, d\mathbf{x} = \int \rho \left( \alpha + \beta u(\mathbf{x}) \right) d\mathbf{x} = \alpha + \beta U
\end{align*}
since $\rho$ satisfies the constraints. However, since $\rho^\ast$ also satisfies the constraints,
\begin{align*}
H(\rho^\ast) = - \int \rho^\ast \ln \rho^\ast \, d\mathbf{x} = \int \rho^\ast \left( \alpha + \beta u(\mathbf{x}) \right) d\mathbf{x} = \alpha + \beta U
\end{align*}
which implies that $H(\rho) \leq H(\rho^\ast)$.
\end{proof}

\section{Bounds on field energy decay D}
\label{sec:sl_opt_details}

To handle the general case of both 2D and 3D focusing, we define $n$ as the number of dimensions in the focal plane, $\mathbf{k}_\perp$ as the generalized transverse wavevector, and $\sigma$ and $\hat{\sigma}$ as the spot size in real space and k-space respectively. In 2D focusing, $n = 1$ and $\mathbf{k}_\perp = (k_x)$. In addition, the spot sizes are widths, i.e. $\sigma = w$ and $\hat{\sigma} = \hat{w}$. Meanwhile, in 3D focusing, $n = 2$, $\mathbf{k}_\perp = (k_x, k_y)$, and the spot sizes are areas, i.e. $\sigma = A$ and $\hat{\sigma} = \hat{A}$. 

As we discuss in the main text, we first find a lower bound on the k-space spot size $\hat{\sigma}$ as a function of the focal distance $L$ and field energy decay $D$ by solving the optimization problem
\begin{alignat}{3}
&\maximize  \quad & \ln \hat{\sigma} &= -\textstyle\int \rho(\mathbf{k}_\perp) \ln \rho(\mathbf{k}_\perp) \; d\mathbf{k}_\perp  \nonumber \\
&\subjectto \quad & 0 &\leq \rho(\mathbf{k}_\perp)  \nonumber \\
&                 & 1 &= \textstyle\int \rho(\mathbf{k}_\perp) \;d\mathbf{k}_\perp \nonumber \\
&                 & \frac{1}{D} &= \textstyle\int \rho(\mathbf{k}_\perp) \; \left| e^{- i k_z L} \right|^2  \; d\mathbf{k}_\perp .
\label{eqn:sl_opt_prob_gen}
\end{alignat}
From Appendix \ref{sec:sl_entropyopt}, the distribution $\rho^\ast(\mathbf{k}_\perp)$ that solves the optimization problem (\ref{eqn:sl_opt_prob_gen}) is
\begin{align}
\rho^\ast(\mathbf{k}_\perp) = N \exp\left(- \beta \left| e^{- i k_z L} \right|^2  \right)
\label{eqn:sl_p_opt}
\end{align}
where $N, \, \beta > 0$. For any given $\beta$, we can find the normalization constant $N$ using
\begin{align}
\frac{1}{N} = \int \exp\left(- \beta \left| e^{- i k_z L} \right|^2  \right) d\mathbf{k}_\perp.
\label{eqn:sl_N_opt}
\end{align}
The k-space spot size $\hat{\sigma}$ and field energy decay $D$ of the optimal distribution $\rho^\ast(\mathbf{k}_\perp)$ are then functions of $\beta$:
\begin{align}
\ln \hat{\sigma} &= - \int \rho^\ast(\mathbf{k}_\perp) \ln \rho^\ast(\mathbf{k}_\perp) \,  d\mathbf{k}_\perp \label{eqn:sl_sk_opt} \\
\frac{1}{D} &= \int \rho^\ast(\mathbf{k}_\perp) \, \left| e^{- i k_z L} \right|^2  \, d\mathbf{k}_\perp.
\label{eqn:sl_D_opt}
\end{align}
By sweeping $\beta$ over the set of all positive numbers, we can obtain any field energy decay $D$. 

In this way, we have found an upper bound for the k-space spot size $\hat{\sigma}$ of the form
\begin{align}
F(D) \geq \hat{\sigma} 
\end{align}
where $F: \mathbb{R} \rightarrow \mathbb{R}$ is a decreasing function of $D$, and the focal distance $L$ is assumed to be fixed. We can map the k-space spot size to an equivalent real-space spot size using the uncertainty principle
\begin{align}
\sigma \, \hat{\sigma} \geq e^{n (1 + \ln \pi)}
\label{eqn:sl_uncertainty_suppl}
\end{align}
to yield
\begin{align}
F(D) \geq \hat{\sigma} \geq \frac{e^{n(\ln \pi + 1)}}{\sigma }.
\label{eqn:sl_D_sk_sr_ineq}
\end{align}
The inverse of $F(D)$ exists on $0 < D < 1$ since $F$ is monotonically decreasing over this domain. We can thus apply $F^{-1}$ to both sides of (\ref{eqn:sl_D_sk_sr_ineq}) to obtain an upper bound on the field energy decay $D$:
\begin{align}
D \leq F^{-1}\left( \frac{e^{n(\ln \pi + 1)}}{\sigma }\right).
\end{align}

\section{Far-field regime}
\label{sec:sl_lim_farfield}
In the far-field regime, the fields in the focal plane can only consist of propagating waves. In k-space, this corresponds to limiting the transverse spatial frequencies to the circle
\begin{align}
k_x^2 + k_y^2 \leq \left(\frac{2 \pi}{\lambda} \right)^2.
\label{eqn:sl_farfield_circle}
\end{align}
For convenience, we will use the notation $k_0 = 2 \pi / \lambda$ for the wavenumber of propagating waves.

\subsection{2D focusing}
In two-dimensional focusing, we have a single tranverse spatial frequency $k_x$. The distribution $\rho^\ast$ that maximizes the k-space spot width $\hat{w}$ while remaining within the circle (\ref{eqn:sl_farfield_circle}) is the uniform distribution
\begin{align}
\rho^\ast(k_x) = 
\frac{1}{2 k_0}
\begin{cases}
1, & |k_x| \leq k_0 \\
0, & \text{otherwise}.
\end{cases}
\end{align}
This distribution has a k-space spot width $\hat{w} = 2 k_0$. The real-space spot width $\hat{w}$ is thus limited by the uncertainty principle (\ref{eqn:sl_uncertainty_suppl}) to
\begin{align}
w \geq \frac{e}{4} \lambda \approx 0.680 \, \lambda.
\end{align}
This corresponds nicely to the commonly used rule of thumb that the resolution of a conventional microscope is limited to approximately half a wavelength.

\subsection{3D focusing}
In three-dimensional focusing, we have two transverse spatial frequencies $k_x$ and $k_y$. The optimal distribution is again a uniform distribution
\begin{align}
\rho^\ast(k_x, k_y) = 
\frac{1}{\pi k_0^2}
\begin{cases}
1, & k_x^2 + k_y^2 \leq k_0^2 \\
0, & \text{otherwise}.
\end{cases}
\end{align}
with a k-space spot area of $\hat{A} = \pi k_0^2$. The real-space spot area $A$ is thus bounded by
\begin{align}
A \geq \frac{e^2}{4 \pi} \lambda^2 \approx 0.588 \, \lambda^2.
\end{align}

\section{Deep subwavelength limit}
\label{sec:sl_lim_subwl}

We will now derive simplified analytic bounds for the field energy decay $D$ in the deep subwavelength regime. As we discussed in the main text, in this regime we can make the approximation
\begin{align}
\left| e^{- i k_z L} \right|^2  \approx e^{2 L | \mathbf{k}_\perp|}.
\end{align}
From (\ref{eqn:sl_p_opt}), the optimal distribution in k-space is then of the form
\begin{align}
\rho^\ast(\mathbf{k}_\perp) = N e^{- \beta e^{2 L |\mathbf{k}_\perp|}}
\end{align}
where $N$ and $\beta$ are positive constants. 

\subsection{2D focusing}
In two dimensional focusing, the number of dimensions in the focal plane $n = 1$, and we have a single transverse spatial frequency $k_x$. The normalization $N$ is then given by (\ref{eqn:sl_N_opt}) as
\begin{align}
\frac{1}{N} = \int_{-\infty}^{\infty} e^{- \beta e^{2 L |k_x|}} dk_x = -\frac{\Ei(-\beta)}{L}
\label{eqn:sl_N_2d}
\end{align}
where $\Ei$ is the exponential integral function
\begin{align}
\Ei(x) = - \int_{-x}^{\infty} \frac{e^{-t}}{t} dt.
\end{align}
The optimal k-space spot size $\hat{w}$ is then given by (\ref{eqn:sl_sk_opt}) as
\begin{align}
\hat{w} = - \frac{\Ei(-\beta)}{L} \exp\left( - \frac{e^{-\beta}}{\Ei(-\beta)} \right).
\label{eqn:sl_sk_2d}
\end{align}
Finally, the field-energy decay $D$ is given by (\ref{eqn:sl_D_opt}) as
\begin{align}
\frac{1}{D} = - \frac{e^{-\beta}}{\beta \Ei(-\beta)} .
\label{eqn:sl_D_2d}
\end{align}

We can simplify these expressions significantly in the \emph{small $\beta$ limit} where $\ln \beta \ll -1$. In this regime, we can make use of the approximations $\exp(x) \approx 1 + x$ and $\Ei(x) \approx \gamma + \ln|x|$ for $|x| \ll 1$. Here, $\gamma \approx 0.577$ is the Euler-Mascheroni constant. Solving for $\beta$ as a function $\hat{w}$ in (\ref{eqn:sl_sk_2d}) then yields
\begin{align}
\ln \beta \approx - \hat{w} L - \gamma + 1.
\label{eqn:sl_beta_2da}
\end{align}
Combining (\ref{eqn:sl_beta_2da}) with the uncertainty principle (\ref{eqn:sl_uncertainty_suppl}), we can see that we are in the small $\beta$ limit if the focusing distance $L$ is sufficiently large such that
\begin{align}
\frac{L}{w} \gg \frac{2 - \gamma}{e \pi} \approx 0.167.
\label{eqn:sl_small_beta_2d}
\end{align}
Next, we can simplify our expression for the field energy decay $D$ by combining (\ref{eqn:sl_D_2d}) with (\ref{eqn:sl_beta_2da}) and again applying our approximations:
\begin{align}
D \approx \frac{\hat{w} L + 1}{\exp\left( \hat{w} L + \gamma - 1\right)}.
\end{align}
Finally, using the uncertainty principle (\ref{eqn:sl_uncertainty_suppl}), we obtain the following upper bound on $D$:
\begin{align}
D \leq \left(\frac{L}{\delta_{2D}} + 1 \right) \exp\left( - \frac{L}{\delta_{2D}} + 1 - \gamma \right),
\label{eqn:sl_constr_2d_approx}
\end{align}
where the decay length $\delta_{2D}$ is proportional to the spot width $w$,
\begin{align}
\delta_{2D} = \frac{w}{\pi e} \approx 0.12 \, w.
\label{eqn:sl_constr_2d_R}
\end{align}
Thus, if the focusing distance is sufficently large that it satisfies (\ref{eqn:sl_small_beta_2d}), the fields decay exponentially from the focusing system to the focal plane.

\subsection{3D focusing}
\begin{figure}[h!tb]
\centering
\includegraphics[scale=0.6]{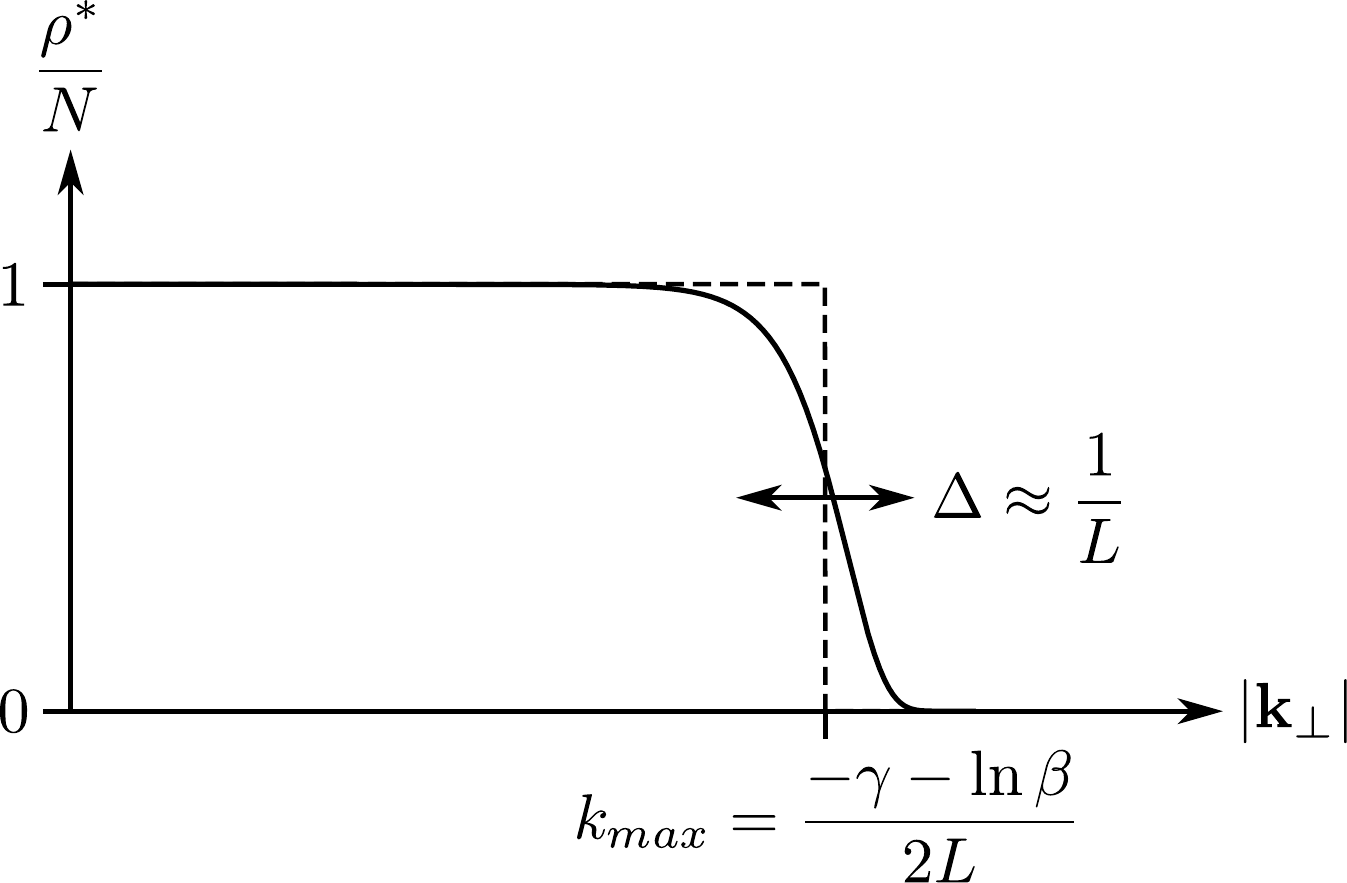}
\caption{Radial distribution of $\rho^\ast(\mathbf{k}_\perp)$ in the 3D case. $\rho^\ast$ is well approximated by a step function when $k_{max} / \Delta \gg 1$, which corresponds to $\ln \beta \ll -1$. }
\label{fig:sl_step_approx}
\end{figure}

In three dimensional focusing, the number of dimensions in the focal plane $n = 2$, and we have a two transverse spatial frequencies $k_x$ and $k_y$. The normalization constant is given by (\ref{eqn:sl_N_opt}) as
\begin{align}
\frac{1}{N} = \iint e^{- \beta e^{2 L |\mathbf{k}_\perp|}} d\mathbf{k}_\perp = 2 \pi \int_{0}^{\infty} e^{- \beta e^{2 L k}} \, k \, dk.
\label{eqn:sl_N_3d}
\end{align}
Unfortunately, (\ref{eqn:sl_N_3d}) has no clean analytic solution. In the \emph{small $\beta$ limit} where $\ln\beta \ll -1$, however, we can approximate $\rho^\ast$ for the purposes of computing $N$ with
\begin{align}
\rho^\ast(\mathbf{k}_\perp) \approx N \, \Theta\left( | \mathbf{k}_\perp | \right) 
\label{eqn:sl_N_3da}
\end{align}
where $\Theta(k)$ is the step function
\begin{align}
\Theta(k) = 
\begin{cases}
1, & k \leq \dfrac{- \gamma - \ln \beta}{2 L} \\
0, & \text{otherwise}
\end{cases}
\end{align}
as illustrated in figure \ref{fig:sl_step_approx}. This particular choice of step function satisfies
\begin{align}
\int_\Omega^{\infty} e^{-\beta e^{2 L k}} \, dk = \int_\Omega^{\infty} \Theta(k) \, dk 
\end{align}
in the limit $\Omega \rightarrow - \infty$. 

With the step-function approximation, the normalization constant $N$ becomes
\begin{align}
\frac{1}{N} \approx \pi \left( \frac{\gamma + \ln \beta}{2 L} \right)^2.
\label{eqn:sl_N_3db}
\end{align}
The k-space spot size $\hat{A}$ is then given by (\ref{eqn:sl_sk_opt}) as
\begin{align}
\hat{A} &= \frac{1}{N} \exp\left( - \frac{\pi N}{2 L^2} \Ei(-\beta) \right) \\
&\approx \pi \left( \frac{\gamma + \ln \beta}{2 L} \right)^2 \exp\left(  -\frac{2 \Ei(-\beta)}{\beta \left( \gamma + \ln \beta \right)^2 } \right).
\label{eqn:sl_sk_3d}
\end{align}
Finally, the field energy decay $D$ is given by (\ref{eqn:sl_D_opt}) as
\begin{align}
\frac{1}{D} = -\frac{\pi N}{2 L^2 \beta} \Ei(-\beta)
\approx - \frac{2 \Ei(-\beta)}{\beta \left( \gamma + \ln \beta\right)^2}.
\label{eqn:sl_D_3d}
\end{align}

We can simplify these expressions in the small $\beta$ limit where $\ln \beta \ll -1$ by using the same approximations as in the one-dimensional case. In this limit, we can solve for $\beta$ in (\ref{eqn:sl_sk_3d}) to yield
\begin{align}
\ln \beta \approx 1 - \gamma - \sqrt{1 + \frac{4 \hat{A} L}{\pi}} .
\label{eqn:sl_beta_3da}
\end{align}
Combining (\ref{eqn:sl_beta_3da}) with the uncertainty principle (\ref{eqn:sl_uncertainty_suppl}), we again find that we are in the small $\beta$ limit if the focusing distance $L$ is sufficiently large, with the slightly modified constraint
\begin{align}
\frac{L}{\sqrt{A}} \gg \frac{1}{2 e \sqrt{\pi}} \approx 0.10.
\label{eqn:sl_small_beta_3d}
\end{align}
Finally, we obtain an upper bound on the field energy decay $D$ by combining (\ref{eqn:sl_D_3d}) with (\ref{eqn:sl_beta_3da}), applying our approximations, and making use of the uncertainty principle (\ref{eqn:sl_uncertainty_suppl}), yielding
\begin{align}
D \leq \frac{L}{2 \, \delta_{3D}} \exp\left(- \frac{L}{\delta_{3D}} - \gamma \right)
\label{eqn:sl_constr_3d_approx}
\end{align}
where the decay length $\delta_{3D}$ is approximately proportional to the square root of the spot area $A$:
\begin{align}
\delta_{3D} = L \left(-1 + \sqrt{1 + \frac{4 \pi e^2 L^2}{A}} \right)^{-1} \approx \frac{\sqrt{A}}{2 \sqrt{\pi} e} \approx 0.10 \sqrt{A}.
\label{eqn:sl_constr_3d_R}
\end{align}
Thus, if the focusing distance $L$ is sufficently large that it satisfies (\ref{eqn:sl_small_beta_3d}), the fields decay exponentially from the focusing system to the focal plane.


\begin{thebibliography}{28}%
\makeatletter
\providecommand \@ifxundefined [1]{%
 \@ifx{#1\undefined}
}%
\providecommand \@ifnum [1]{%
 \ifnum #1\expandafter \@firstoftwo
 \else \expandafter \@secondoftwo
 \fi
}%
\providecommand \@ifx [1]{%
 \ifx #1\expandafter \@firstoftwo
 \else \expandafter \@secondoftwo
 \fi
}%
\providecommand \natexlab [1]{#1}%
\providecommand \enquote  [1]{``#1''}%
\providecommand \bibnamefont  [1]{#1}%
\providecommand \bibfnamefont [1]{#1}%
\providecommand \citenamefont [1]{#1}%
\providecommand \href@noop [0]{\@secondoftwo}%
\providecommand \href [0]{\begingroup \@sanitize@url \@href}%
\providecommand \@href[1]{\@@startlink{#1}\@@href}%
\providecommand \@@href[1]{\endgroup#1\@@endlink}%
\providecommand \@sanitize@url [0]{\catcode `\\12\catcode `\$12\catcode
  `\&12\catcode `\#12\catcode `\^12\catcode `\_12\catcode `\%12\relax}%
\providecommand \@@startlink[1]{}%
\providecommand \@@endlink[0]{}%
\providecommand \url  [0]{\begingroup\@sanitize@url \@url }%
\providecommand \@url [1]{\endgroup\@href {#1}{\urlprefix }}%
\providecommand \urlprefix  [0]{URL }%
\providecommand \Eprint [0]{\href }%
\providecommand \doibase [0]{http://dx.doi.org/}%
\providecommand \selectlanguage [0]{\@gobble}%
\providecommand \bibinfo  [0]{\@secondoftwo}%
\providecommand \bibfield  [0]{\@secondoftwo}%
\providecommand \translation [1]{[#1]}%
\providecommand \BibitemOpen [0]{}%
\providecommand \bibitemStop [0]{}%
\providecommand \bibitemNoStop [0]{.\EOS\space}%
\providecommand \EOS [0]{\spacefactor3000\relax}%
\providecommand \BibitemShut  [1]{\csname bibitem#1\endcsname}%
\let\auto@bib@innerbib\@empty
%</preamble>
\bibitem [{\citenamefont {Dunn}(1999)}]{rcdunn_cr1999}%
  \BibitemOpen
  \bibfield  {author} {\bibinfo {author} {\bibfnamefont {R.~C.}\ \bibnamefont
  {Dunn}},\ }\href {\doibase 10.1021/cr980130e} {\bibfield  {journal} {\bibinfo
   {journal} {Chem. Rev.}\ }\textbf {\bibinfo {volume} {99}},\ \bibinfo {pages}
  {2891} (\bibinfo {year} {1999})},\ \bibinfo {note} {pMID:
  11749505}\BibitemShut {NoStop}%
\bibitem [{\citenamefont {Pendry}(2000)}]{jbpendry_prl2000}%
  \BibitemOpen
  \bibfield  {author} {\bibinfo {author} {\bibfnamefont {J.~B.}\ \bibnamefont
  {Pendry}},\ }\href {\doibase 10.1103/PhysRevLett.85.3966} {\bibfield
  {journal} {\bibinfo  {journal} {Phys. Rev. Lett.}\ }\textbf {\bibinfo
  {volume} {85}},\ \bibinfo {pages} {3966} (\bibinfo {year}
  {2000})}\BibitemShut {NoStop}%
\bibitem [{\citenamefont {Fang}\ \emph {et~al.}(2005)\citenamefont {Fang},
  \citenamefont {Lee}, \citenamefont {Sun},\ and\ \citenamefont
  {Zhang}}]{nfang_science2005}%
  \BibitemOpen
  \bibfield  {author} {\bibinfo {author} {\bibfnamefont {N.}~\bibnamefont
  {Fang}}, \bibinfo {author} {\bibfnamefont {H.}~\bibnamefont {Lee}}, \bibinfo
  {author} {\bibfnamefont {C.}~\bibnamefont {Sun}}, \ and\ \bibinfo {author}
  {\bibfnamefont {X.}~\bibnamefont {Zhang}},\ }\href {\doibase
  10.1126/science.1108759} {\bibfield  {journal} {\bibinfo  {journal}
  {Science}\ }\textbf {\bibinfo {volume} {308}},\ \bibinfo {pages} {534}
  (\bibinfo {year} {2005})}\BibitemShut {NoStop}%
\bibitem [{\citenamefont {Zhang}\ and\ \citenamefont
  {Liu}(2008)}]{xzhang_nmat2008}%
  \BibitemOpen
  \bibfield  {author} {\bibinfo {author} {\bibfnamefont {X.}~\bibnamefont
  {Zhang}}\ and\ \bibinfo {author} {\bibfnamefont {Z.}~\bibnamefont {Liu}},\
  }\href {\doibase 10.1038/nmat2141} {\bibfield  {journal} {\bibinfo  {journal}
  {Nat. Mater.}\ }\textbf {\bibinfo {volume} {7}},\ \bibinfo {pages} {435}
  (\bibinfo {year} {2008})}\BibitemShut {NoStop}%
\bibitem [{\citenamefont {Poddubny}\ \emph {et~al.}(2013)\citenamefont
  {Poddubny}, \citenamefont {Iorsh}, \citenamefont {Belov},\ and\ \citenamefont
  {Kivshar}}]{apoddubny_np2013}%
  \BibitemOpen
  \bibfield  {author} {\bibinfo {author} {\bibfnamefont {A.}~\bibnamefont
  {Poddubny}}, \bibinfo {author} {\bibfnamefont {I.}~\bibnamefont {Iorsh}},
  \bibinfo {author} {\bibfnamefont {P.}~\bibnamefont {Belov}}, \ and\ \bibinfo
  {author} {\bibfnamefont {Y.}~\bibnamefont {Kivshar}},\ }\href@noop {}
  {\bibfield  {journal} {\bibinfo  {journal} {Nat. Photon.}\ }\textbf {\bibinfo
  {volume} {7}},\ \bibinfo {pages} {948} (\bibinfo {year} {2013})},\ \bibinfo
  {note} {review}\BibitemShut {NoStop}%
\bibitem [{\citenamefont {Liu}\ \emph {et~al.}(2007)\citenamefont {Liu},
  \citenamefont {Lee}, \citenamefont {Xiong}, \citenamefont {Sun},\ and\
  \citenamefont {Zhang}}]{zliu_science2007}%
  \BibitemOpen
  \bibfield  {author} {\bibinfo {author} {\bibfnamefont {Z.}~\bibnamefont
  {Liu}}, \bibinfo {author} {\bibfnamefont {H.}~\bibnamefont {Lee}}, \bibinfo
  {author} {\bibfnamefont {Y.}~\bibnamefont {Xiong}}, \bibinfo {author}
  {\bibfnamefont {C.}~\bibnamefont {Sun}}, \ and\ \bibinfo {author}
  {\bibfnamefont {X.}~\bibnamefont {Zhang}},\ }\href {\doibase
  10.1126/science.1137368} {\bibfield  {journal} {\bibinfo  {journal}
  {Science}\ }\textbf {\bibinfo {volume} {315}},\ \bibinfo {pages} {1686}
  (\bibinfo {year} {2007})}\BibitemShut {NoStop}%
\bibitem [{\citenamefont {Lu}\ and\ \citenamefont {Liu}(2012)}]{dlu_ncomm2012}%
  \BibitemOpen
  \bibfield  {author} {\bibinfo {author} {\bibfnamefont {D.}~\bibnamefont
  {Lu}}\ and\ \bibinfo {author} {\bibfnamefont {Z.}~\bibnamefont {Liu}},\
  }\href@noop {} {\bibfield  {journal} {\bibinfo  {journal} {Nat. Commun.}\
  }\textbf {\bibinfo {volume} {3}},\ \bibinfo {pages} {1205} (\bibinfo {year}
  {2012})}\BibitemShut {NoStop}%
\bibitem [{\citenamefont {Dai}\ \emph {et~al.}(2015)\citenamefont {Dai},
  \citenamefont {Ma}, \citenamefont {Andersen}, \citenamefont {Mcleod},
  \citenamefont {Fei}, \citenamefont {Liu}, \citenamefont {Wagner},
  \citenamefont {Watanabe}, \citenamefont {Taniguchi}, \citenamefont
  {Thiemens}, \citenamefont {Keilmann}, \citenamefont {Jarillo-Herrero},
  \citenamefont {Fogler},\ and\ \citenamefont {Basov}}]{sdai_ncomm2015}%
  \BibitemOpen
  \bibfield  {author} {\bibinfo {author} {\bibfnamefont {S.}~\bibnamefont
  {Dai}}, \bibinfo {author} {\bibfnamefont {Q.}~\bibnamefont {Ma}}, \bibinfo
  {author} {\bibfnamefont {T.}~\bibnamefont {Andersen}}, \bibinfo {author}
  {\bibfnamefont {A.~S.}\ \bibnamefont {Mcleod}}, \bibinfo {author}
  {\bibfnamefont {Z.}~\bibnamefont {Fei}}, \bibinfo {author} {\bibfnamefont
  {M.~K.}\ \bibnamefont {Liu}}, \bibinfo {author} {\bibfnamefont
  {M.}~\bibnamefont {Wagner}}, \bibinfo {author} {\bibfnamefont
  {K.}~\bibnamefont {Watanabe}}, \bibinfo {author} {\bibfnamefont
  {T.}~\bibnamefont {Taniguchi}}, \bibinfo {author} {\bibfnamefont
  {M.}~\bibnamefont {Thiemens}}, \bibinfo {author} {\bibfnamefont
  {F.}~\bibnamefont {Keilmann}}, \bibinfo {author} {\bibfnamefont
  {P.}~\bibnamefont {Jarillo-Herrero}}, \bibinfo {author} {\bibfnamefont
  {M.~M.}\ \bibnamefont {Fogler}}, \ and\ \bibinfo {author} {\bibfnamefont
  {D.~N.}\ \bibnamefont {Basov}},\ }\href {\doibase 10.1038/ncomms7963}
  {\bibfield  {journal} {\bibinfo  {journal} {Nat. Commun.}\ }\textbf {\bibinfo
  {volume} {6}},\ \bibinfo {pages} {6963} (\bibinfo {year} {2015})}\BibitemShut
  {NoStop}%
\bibitem [{\citenamefont {Leviatan}(1986)}]{yleviatan_jap1986}%
  \BibitemOpen
  \bibfield  {author} {\bibinfo {author} {\bibfnamefont {Y.}~\bibnamefont
  {Leviatan}},\ }\href {\doibase 10.1063/1.337294} {\bibfield  {journal}
  {\bibinfo  {journal} {Journal of Applied Physics}\ }\textbf {\bibinfo
  {volume} {60}},\ \bibinfo {pages} {1577} (\bibinfo {year}
  {1986})}\BibitemShut {NoStop}%
\bibitem [{\citenamefont {Vigoureux}\ and\ \citenamefont
  {Courjon}(1992)}]{jmvigoureux_ao1992i}%
  \BibitemOpen
  \bibfield  {author} {\bibinfo {author} {\bibfnamefont {J.~M.}\ \bibnamefont
  {Vigoureux}}\ and\ \bibinfo {author} {\bibfnamefont {D.}~\bibnamefont
  {Courjon}},\ }\href {\doibase 10.1364/AO.31.003170} {\bibfield  {journal}
  {\bibinfo  {journal} {Appl. Opt.}\ }\textbf {\bibinfo {volume} {31}},\
  \bibinfo {pages} {3170} (\bibinfo {year} {1992})}\BibitemShut {NoStop}%
\bibitem [{\citenamefont {Vigoureux}\ \emph {et~al.}(1992)\citenamefont
  {Vigoureux}, \citenamefont {Depasse},\ and\ \citenamefont
  {Girard}}]{jmvigoureux_ao1992ii}%
  \BibitemOpen
  \bibfield  {author} {\bibinfo {author} {\bibfnamefont {J.~M.}\ \bibnamefont
  {Vigoureux}}, \bibinfo {author} {\bibfnamefont {F.}~\bibnamefont {Depasse}},
  \ and\ \bibinfo {author} {\bibfnamefont {C.}~\bibnamefont {Girard}},\ }\href
  {\doibase 10.1364/AO.31.003036} {\bibfield  {journal} {\bibinfo  {journal}
  {Appl. Opt.}\ }\textbf {\bibinfo {volume} {31}},\ \bibinfo {pages} {3036}
  (\bibinfo {year} {1992})}\BibitemShut {NoStop}%
\bibitem [{\citenamefont {Novotny}\ \emph {et~al.}(1998)\citenamefont
  {Novotny}, \citenamefont {Hecht},\ and\ \citenamefont
  {Pohl}}]{lnovotny_um1998}%
  \BibitemOpen
  \bibfield  {author} {\bibinfo {author} {\bibfnamefont {L.}~\bibnamefont
  {Novotny}}, \bibinfo {author} {\bibfnamefont {B.}~\bibnamefont {Hecht}}, \
  and\ \bibinfo {author} {\bibfnamefont {D.~W.}\ \bibnamefont {Pohl}},\ }\href
  {\doibase https://doi.org/10.1016/S0304-3991(97)00066-1} {\bibfield
  {journal} {\bibinfo  {journal} {Ultramicroscopy}\ }\textbf {\bibinfo {volume}
  {71}},\ \bibinfo {pages} {341 } (\bibinfo {year} {1998})}\BibitemShut
  {NoStop}%
\bibitem [{\citenamefont {Merlin}(2007)}]{rmerlin_science2007}%
  \BibitemOpen
  \bibfield  {author} {\bibinfo {author} {\bibfnamefont {R.}~\bibnamefont
  {Merlin}},\ }\href {\doibase 10.1126/science.1143884} {\bibfield  {journal}
  {\bibinfo  {journal} {Science}\ }\textbf {\bibinfo {volume} {317}},\ \bibinfo
  {pages} {927} (\bibinfo {year} {2007})}\BibitemShut {NoStop}%
\bibitem [{\citenamefont {Intaraprasonk}\ and\ \citenamefont
  {Fan}(2009)}]{vintaraprasonk_ol2009}%
  \BibitemOpen
  \bibfield  {author} {\bibinfo {author} {\bibfnamefont {V.}~\bibnamefont
  {Intaraprasonk}}\ and\ \bibinfo {author} {\bibfnamefont {S.}~\bibnamefont
  {Fan}},\ }\href {\doibase 10.1364/OL.34.002967} {\bibfield  {journal}
  {\bibinfo  {journal} {Opt. Lett.}\ }\textbf {\bibinfo {volume} {34}},\
  \bibinfo {pages} {2967} (\bibinfo {year} {2009})}\BibitemShut {NoStop}%
\bibitem [{\citenamefont {Grbic}\ \emph {et~al.}(2008)\citenamefont {Grbic},
  \citenamefont {Jiang},\ and\ \citenamefont {Merlin}}]{agrbic_science2008}%
  \BibitemOpen
  \bibfield  {author} {\bibinfo {author} {\bibfnamefont {A.}~\bibnamefont
  {Grbic}}, \bibinfo {author} {\bibfnamefont {L.}~\bibnamefont {Jiang}}, \ and\
  \bibinfo {author} {\bibfnamefont {R.}~\bibnamefont {Merlin}},\ }\href
  {\doibase 10.1126/science.1154753} {\bibfield  {journal} {\bibinfo  {journal}
  {Science}\ }\textbf {\bibinfo {volume} {320}},\ \bibinfo {pages} {511}
  (\bibinfo {year} {2008})}\BibitemShut {NoStop}%
\bibitem [{\citenamefont {Lenzner}\ \emph {et~al.}(1998)\citenamefont
  {Lenzner}, \citenamefont {Kr\"uger}, \citenamefont {Sartania}, \citenamefont
  {Cheng}, \citenamefont {Spielmann}, \citenamefont {Mourou}, \citenamefont
  {Kautek},\ and\ \citenamefont {Krausz}}]{mlenzer_prl1998}%
  \BibitemOpen
  \bibfield  {author} {\bibinfo {author} {\bibfnamefont {M.}~\bibnamefont
  {Lenzner}}, \bibinfo {author} {\bibfnamefont {J.}~\bibnamefont {Kr\"uger}},
  \bibinfo {author} {\bibfnamefont {S.}~\bibnamefont {Sartania}}, \bibinfo
  {author} {\bibfnamefont {Z.}~\bibnamefont {Cheng}}, \bibinfo {author}
  {\bibfnamefont {C.}~\bibnamefont {Spielmann}}, \bibinfo {author}
  {\bibfnamefont {G.}~\bibnamefont {Mourou}}, \bibinfo {author} {\bibfnamefont
  {W.}~\bibnamefont {Kautek}}, \ and\ \bibinfo {author} {\bibfnamefont
  {F.}~\bibnamefont {Krausz}},\ }\href {\doibase 10.1103/PhysRevLett.80.4076}
  {\bibfield  {journal} {\bibinfo  {journal} {Phys. Rev. Lett.}\ }\textbf
  {\bibinfo {volume} {80}},\ \bibinfo {pages} {4076} (\bibinfo {year}
  {1998})}\BibitemShut {NoStop}%
\bibitem [{\citenamefont {{Supplementary Material}}()}]{suppl}%
  \BibitemOpen
  \bibfield  {author} {\bibinfo {author} {\bibnamefont {{Supplementary
  Material}}},\ }\href@noop {} {\ }\BibitemShut {NoStop}%
\bibitem [{\citenamefont {Harrington}(2001)}]{rfharrington_2001_helmholtz}%
  \BibitemOpen
  \bibfield  {author} {\bibinfo {author} {\bibfnamefont {R.~F.}\ \bibnamefont
  {Harrington}},\ }\enquote {\bibinfo {title} {Time-harmonic electromagnetic
  fields},}\ \ (\bibinfo  {publisher} {IEEE Press},\ \bibinfo {year} {2001})\
  pp.\ \bibinfo {pages} {37--38}\BibitemShut {NoStop}%
\bibitem [{\citenamefont {Porras}(1994)}]{maporras_oc1994}%
  \BibitemOpen
  \bibfield  {author} {\bibinfo {author} {\bibfnamefont {M.~A.}\ \bibnamefont
  {Porras}},\ }\href {\doibase 10.1016/0030-4018(94)90729-3} {\bibfield
  {journal} {\bibinfo  {journal} {Optics communications}\ }\textbf {\bibinfo
  {volume} {109}},\ \bibinfo {pages} {5} (\bibinfo {year} {1994})}\BibitemShut
  {NoStop}%
\bibitem [{\citenamefont {Hyv\"{a}rinen}(1997)}]{ahyvarinen_1997}%
  \BibitemOpen
  \bibfield  {author} {\bibinfo {author} {\bibfnamefont {A.}~\bibnamefont
  {Hyv\"{a}rinen}},\ }\href@noop {} {\ ,\ \bibinfo {pages} {273} (\bibinfo
  {year} {1997})}\BibitemShut {NoStop}%
\bibitem [{\citenamefont {Goldie}\ and\ \citenamefont
  {Pinch}(1991)}]{cmgoldie_1991}%
  \BibitemOpen
  \bibfield  {author} {\bibinfo {author} {\bibfnamefont {C.~M.}\ \bibnamefont
  {Goldie}}\ and\ \bibinfo {author} {\bibfnamefont {R.~G.~E.}\ \bibnamefont
  {Pinch}},\ }\href@noop {} {\emph {\bibinfo {title} {Communication Theory}}},\
  London Mathematical Society St\ (\bibinfo  {publisher} {Cambridge University
  Press},\ \bibinfo {year} {1991})\BibitemShut {NoStop}%
\bibitem [{\citenamefont {Lazo}\ and\ \citenamefont
  {Rathie}(1978)}]{avlazo_ieeetoit1978}%
  \BibitemOpen
  \bibfield  {author} {\bibinfo {author} {\bibfnamefont {A.~V.}\ \bibnamefont
  {Lazo}}\ and\ \bibinfo {author} {\bibfnamefont {P.}~\bibnamefont {Rathie}},\
  }\href {\doibase 10.1109/TIT.1978.1055832} {\bibfield  {journal} {\bibinfo
  {journal} {IEEE Trans. Inf. Theory}\ }\textbf {\bibinfo {volume} {24}},\
  \bibinfo {pages} {120} (\bibinfo {year} {1978})}\BibitemShut {NoStop}%
\bibitem [{\citenamefont {Porras}\ and\ \citenamefont
  {Medina}(1995)}]{maporras_ao1995}%
  \BibitemOpen
  \bibfield  {author} {\bibinfo {author} {\bibfnamefont {M.~A.}\ \bibnamefont
  {Porras}}\ and\ \bibinfo {author} {\bibfnamefont {R.}~\bibnamefont
  {Medina}},\ }\href {\doibase 10.1364/AO.34.008247} {\bibfield  {journal}
  {\bibinfo  {journal} {Appl. Opt.}\ }\textbf {\bibinfo {volume} {34}},\
  \bibinfo {pages} {8247} (\bibinfo {year} {1995})}\BibitemShut {NoStop}%
\bibitem [{\citenamefont {Bialynicki-Birula}\ and\ \citenamefont
  {Rudnicki}(2011)}]{ibialynickibirula_2011}%
  \BibitemOpen
  \bibfield  {author} {\bibinfo {author} {\bibfnamefont {I.}~\bibnamefont
  {Bialynicki-Birula}}\ and\ \bibinfo {author} {\bibfnamefont
  {{\L}.}~\bibnamefont {Rudnicki}},\ }in\ \href {\doibase
  10.1007/978-90-481-3890-6_1} {\emph {\bibinfo {booktitle} {Statistical
  Complexity}}}\ (\bibinfo  {publisher} {Springer},\ \bibinfo {year} {2011})\
  pp.\ \bibinfo {pages} {1--34}\BibitemShut {NoStop}%
\bibitem [{\citenamefont {Wehner}\ and\ \citenamefont
  {Winter}(2010)}]{swehner_njp2010}%
  \BibitemOpen
  \bibfield  {author} {\bibinfo {author} {\bibfnamefont {S.}~\bibnamefont
  {Wehner}}\ and\ \bibinfo {author} {\bibfnamefont {A.}~\bibnamefont
  {Winter}},\ }\href {\doibase 10.1088/1367-2630/12/2/025009} {\bibfield
  {journal} {\bibinfo  {journal} {New Journal of Physics}\ }\textbf {\bibinfo
  {volume} {12}},\ \bibinfo {pages} {025009} (\bibinfo {year}
  {2010})}\BibitemShut {NoStop}%
\bibitem [{\citenamefont {Coles}\ \emph {et~al.}(2017)\citenamefont {Coles},
  \citenamefont {Berta}, \citenamefont {Tomamichel},\ and\ \citenamefont
  {Wehner}}]{jpcoles_rmp2017}%
  \BibitemOpen
  \bibfield  {author} {\bibinfo {author} {\bibfnamefont {P.~J.}\ \bibnamefont
  {Coles}}, \bibinfo {author} {\bibfnamefont {M.}~\bibnamefont {Berta}},
  \bibinfo {author} {\bibfnamefont {M.}~\bibnamefont {Tomamichel}}, \ and\
  \bibinfo {author} {\bibfnamefont {S.}~\bibnamefont {Wehner}},\ }\href
  {\doibase 10.1103/RevModPhys.89.015002} {\bibfield  {journal} {\bibinfo
  {journal} {Rev. Mod. Phys.}\ }\textbf {\bibinfo {volume} {89}},\ \bibinfo
  {pages} {015002} (\bibinfo {year} {2017})}\BibitemShut {NoStop}%
\bibitem [{\citenamefont {Bia{\l}ynicki-Birula}\ and\ \citenamefont
  {Mycielski}(1975)}]{ibialyniki-birula_cmp1975}%
  \BibitemOpen
  \bibfield  {author} {\bibinfo {author} {\bibfnamefont {I.}~\bibnamefont
  {Bia{\l}ynicki-Birula}}\ and\ \bibinfo {author} {\bibfnamefont
  {J.}~\bibnamefont {Mycielski}},\ }\href {\doibase 10.1007/BF01608825}
  {\bibfield  {journal} {\bibinfo  {journal} {Commun. Math. Phys.}\ }\textbf
  {\bibinfo {volume} {44}},\ \bibinfo {pages} {129} (\bibinfo {year}
  {1975})}\BibitemShut {NoStop}%
\bibitem [{\citenamefont {Hall}(1999)}]{mjwhall_pra1999}%
  \BibitemOpen
  \bibfield  {author} {\bibinfo {author} {\bibfnamefont {M.~J.~W.}\
  \bibnamefont {Hall}},\ }\href {\doibase 10.1103/PhysRevA.59.2602} {\bibfield
  {journal} {\bibinfo  {journal} {Phys. Rev. A}\ }\textbf {\bibinfo {volume}
  {59}},\ \bibinfo {pages} {2602} (\bibinfo {year} {1999})}\BibitemShut
  {NoStop}%
\end{thebibliography}

\begin{thebibliography}{5}%
\makeatletter
\providecommand \@ifxundefined [1]{%
 \@ifx{#1\undefined}
}%
\providecommand \@ifnum [1]{%
 \ifnum #1\expandafter \@firstoftwo
 \else \expandafter \@secondoftwo
 \fi
}%
\providecommand \@ifx [1]{%
 \ifx #1\expandafter \@firstoftwo
 \else \expandafter \@secondoftwo
 \fi
}%
\providecommand \natexlab [1]{#1}%
\providecommand \enquote  [1]{``#1''}%
\providecommand \bibnamefont  [1]{#1}%
\providecommand \bibfnamefont [1]{#1}%
\providecommand \citenamefont [1]{#1}%
\providecommand \href@noop [0]{\@secondoftwo}%
\providecommand \href [0]{\begingroup \@sanitize@url \@href}%
\providecommand \@href[1]{\@@startlink{#1}\@@href}%
\providecommand \@@href[1]{\endgroup#1\@@endlink}%
\providecommand \@sanitize@url [0]{\catcode `\\12\catcode `\$12\catcode
  `\&12\catcode `\#12\catcode `\^12\catcode `\_12\catcode `\%12\relax}%
\providecommand \@@startlink[1]{}%
\providecommand \@@endlink[0]{}%
\providecommand \url  [0]{\begingroup\@sanitize@url \@url }%
\providecommand \@url [1]{\endgroup\@href {#1}{\urlprefix }}%
\providecommand \urlprefix  [0]{URL }%
\providecommand \Eprint [0]{\href }%
\providecommand \doibase [0]{http://dx.doi.org/}%
\providecommand \selectlanguage [0]{\@gobble}%
\providecommand \bibinfo  [0]{\@secondoftwo}%
\providecommand \bibfield  [0]{\@secondoftwo}%
\providecommand \translation [1]{[#1]}%
\providecommand \BibitemOpen [0]{}%
\providecommand \bibitemStop [0]{}%
\providecommand \bibitemNoStop [0]{.\EOS\space}%
\providecommand \EOS [0]{\spacefactor3000\relax}%
\providecommand \BibitemShut  [1]{\csname bibitem#1\endcsname}%
\let\auto@bib@innerbib\@empty
%</preamble>
\bibitem [{\citenamefont {Harrington}(2001)}]{rfharrington_2001}%
  \BibitemOpen
  \bibfield  {author} {\bibinfo {author} {\bibfnamefont {R.~F.}\ \bibnamefont
  {Harrington}},\ }\href@noop {} {\emph {\bibinfo {title} {Time-Harmonic
  Electromagnetic Fields}}}\ (\bibinfo  {publisher} {IEEE Press},\ \bibinfo
  {year} {2001})\BibitemShut {NoStop}%
\bibitem [{\citenamefont {Collin}(1991)}]{recollin_1991}%
  \BibitemOpen
  \bibfield  {author} {\bibinfo {author} {\bibfnamefont {R.~E.}\ \bibnamefont
  {Collin}},\ }\href@noop {} {\emph {\bibinfo {title} {Field Theory of Guided
  Waves}}}\ (\bibinfo  {publisher} {IEEE Press},\ \bibinfo {year}
  {1991})\BibitemShut {NoStop}%
\bibitem [{\citenamefont {Lassas}\ \emph {et~al.}(1998)\citenamefont {Lassas},
  \citenamefont {Cheney},\ and\ \citenamefont {Uhlmann}}]{mlassas_ip1998}%
  \BibitemOpen
  \bibfield  {author} {\bibinfo {author} {\bibfnamefont {M.}~\bibnamefont
  {Lassas}}, \bibinfo {author} {\bibfnamefont {M.}~\bibnamefont {Cheney}}, \
  and\ \bibinfo {author} {\bibfnamefont {G.}~\bibnamefont {Uhlmann}},\ }\href
  {\doibase 10.1088/0266-5611/14/3/017} {\bibfield  {journal} {\bibinfo
  {journal} {Inverse Problems}\ }\textbf {\bibinfo {volume} {14}},\ \bibinfo
  {pages} {679} (\bibinfo {year} {1998})}\BibitemShut {NoStop}%
\bibitem [{\citenamefont {Cheney}\ and\ \citenamefont
  {Isaacson}(1995)}]{mcheney_ip1995}%
  \BibitemOpen
  \bibfield  {author} {\bibinfo {author} {\bibfnamefont {M.}~\bibnamefont
  {Cheney}}\ and\ \bibinfo {author} {\bibfnamefont {D.}~\bibnamefont
  {Isaacson}},\ }\href {\doibase 10.1088/0266-5611/11/4/015} {\bibfield
  {journal} {\bibinfo  {journal} {Inverse Problems}\ }\textbf {\bibinfo
  {volume} {11}},\ \bibinfo {pages} {865} (\bibinfo {year} {1995})}\BibitemShut
  {NoStop}%
\bibitem [{\citenamefont {Goldie}\ and\ \citenamefont
  {Pinch}(1991)}]{cmgoldie_1991}%
  \BibitemOpen
  \bibfield  {author} {\bibinfo {author} {\bibfnamefont {C.~M.}\ \bibnamefont
  {Goldie}}\ and\ \bibinfo {author} {\bibfnamefont {R.~G.~E.}\ \bibnamefont
  {Pinch}},\ }\href@noop {} {\emph {\bibinfo {title} {Communication Theory}}},\
  London Mathematical Society St\ (\bibinfo  {publisher} {Cambridge University
  Press},\ \bibinfo {year} {1991})\BibitemShut {NoStop}%
\end{thebibliography}
\end{document}